\documentclass[11pt]{article}
\usepackage{amssymb,latexsym}
\usepackage{graphicx}
\usepackage{amssymb, amsmath, amsopn, amsthm,slashed}

\usepackage[all]{xy}
\renewcommand{\theequation}{\thesection.\arabic{equation}}
\newcounter{saveeqn}
\newcommand{\add}{\addtocounter{equation}{1}}
\newcommand{\alpheqn}{\setcounter{saveeqn}{\value{equation}}%
\setcounter{equation}{0}%
\renewcommand{\theequation}{\mbox{\thesection.\arabic{saveeqn}{\alph{equation}}}}}
\newcommand{\reseteqn}{\setcounter{equation}{\value{saveeqn}}%
\renewcommand{\theequation}{\thesection.\arabic{equation}}}

\newif\iffigs\figstrue

\DeclareFontFamily{U}{rsf}{}
\DeclareFontShape{U}{rsf}{m}{n}{
  <5> <6> rsfs5 <7> <8> <9> rsfs7 <10-> rsfs10}{}
\DeclareMathAlphabet\Scr{U}{rsf}{m}{n}

\def\pplogo{\vbox{\kern-\headheight\kern -29pt
\halign{##&##\hfil\cr&{
\ppnumber}\cr\rule{0pt}{2.5ex}&\ppdate\cr}
}}
\makeatletter
\def\ps@firstpage{\ps@empty \def\@oddhead{\hss\pplogo}%
  \let\@evenhead\@oddhead 
}
\def\maketitle{\par
 \begingroup
 \def\thefootnote{\fnsymbol{footnote}}
 \def\@makefnmark{\hbox{$^{\@thefnmark}$\hss}}
 \if@twocolumn
 \twocolumn[\@maketitle]
 \else \newpage
 \global\@topnum\z@ \@maketitle \fi\thispagestyle{firstpage}\@thanks
 \endgroup
 \setcounter{footnote}{0}
 \let\maketitle\relax
 \let\@maketitle\relax
 \gdef\@thanks{}\gdef\@author{}\gdef\@title{}\let\thanks\relax}
\makeatother

\newcommand{\bea}{\begin{eqnarray}}
\newcommand{\eea}{\end{eqnarray}}
\newcommand{\be}{\begin{equation}}
\newcommand{\ee}{\end{equation}}

\def\P{{\mathbb P}}

\def\Z{{\mathbb Z}}

\def\IP{{\mathbb P}}

\def\be{\begin{equation}}
\def\beq{\begin{equation}}
\def\ee{\end{equation}}
\def\eeq{\end{equation}}
\def\ba{\begin{eqnarray}}
\def\ea{\end{eqnarray}}
\def\eqn#1{\begin{equation}\begin{split}#1\end{split}\end{equation}}

\def\hd{{\hat d}}

\begin{document}

\thispagestyle{empty}

\begin{titlepage}
\begin{flushright}
LMU-ASC 46/10
\end{flushright}


\vskip  2 cm

\vspace{24pt}

\begin{center}
{ \Large \textbf{ D-brane Superpotentials: Geometric and Worldsheet Approaches}}

\vspace{28pt}

{\bf Marco Baumgartl}\footnote{\normalsize{\ m.baumgartl@physik.uni-muenchen.de}} , {\bf  Ilka Brunner}\footnote{\normalsize{\ ilka.brunner@physik.uni-muenchen.de}} , and \hspace{0.2cm}{\bf
Masoud Soroush}\footnote{\normalsize{\ masoud.soroush@physik.uni-muenchen.de}}

    \vspace{18pt}

 \textsl{Arnold Sommerfeld Center, Ludwig-Maximilians-Universit\"{a}t
    }\\ \textsl{Theresienstr. 37, M\"{u}nchen 80333, Germany}\\
    \vspace{8pt}
 \textsl{Excellence Cluster Universe, Technische Universit\"{a}t M\"{u}nchen}\\
 \textsl{Boltzmannstr. 2, Garching 85748, Germany}

\vspace{10pt}

\vspace{24pt}

\end{center}

\begin{abstract}
From the worldsheet perspective, the superpotential on a D-brane wrapping internal cycles of a Calabi-Yau manifold is given as a generating functional for disk correlation functions.  On the other hand, from the geometric point of view, D-brane superpotentials are captured by certain chain integrals. In this work, we explicitly show for branes wrapping internal 2-cycles how these two different approaches are related. More specifically, from the worldsheet point of view, D-branes at the Landau-Ginzburg point have a convenient description in terms of matrix factorizations. We use a formula derived by Kapustin and Li to explicitly evaluate disk correlators for families of D2-branes. On the geometry side, we then construct a three-chain whose period gives rise to the effective superpotential and show that the two expressions coincide. Finally, as an explicit example, we choose a particular compact Calabi-Yau hypersurface and compute the effective D2-brane superpotential in different branches of the open moduli space, in both geometric and worldsheet approaches.

\end{abstract}

\end{titlepage}
\newpage

\section{Introduction}
\setcounter{equation}{0}
Compactifications of string theory to four dimensions with minimal $N=1$ supersymmetry are of particular interest both from a phenomenological and formal point of view. A particularly important quantity to calculate for this class of string models is the superpotential. Phenomenologically, it contains relevant physical information such as masses and Yukawa couplings of the low energy physics. Mathematically, the superpotential is a holomorphic object, and the canonical quantity to calculate in the context of $N=1$ supersymmetry.

One way to achieve $N=1$ supersymmetry in 4 dimensions is to consider type II string theory with D-branes extending in the uncompactified dimensions. For consistency, one then also needs to include orientifold planes for tadpole cancellation, however, we will not discuss this in the current paper and instead restrict to the D-brane physics. More precisely, we will focus on D-branes wrapping internal 2-cycles, and we will refer to them as D2-branes -- although we have in mind that the configuration could become part of a proper string compactification, where branes extend in the uncompactified directions. The superpotential is then part of a suitably twisted open-closed topological string theory. In particular, if  the brane is B-type, as is the case here, it will depend only on the complex structure  moduli; those are part of the topological sector. On the other hand, it will decouple from  the K\"ahler  moduli of the bulk.

As is well known, mirror symmetry in the bulk connects the B-model on a Calabi-Yau manifold with the A-model on the mirror manifold. An extension to the open string sector states that the category of topological B-branes of the B-model is equivalent to the topological A-branes of the A-model. In particular, on the A-side, the superpotential depends on the K\"ahler moduli and can receive non-perturbative corrections from disk-instantons. On the other hand, the superpotential on a B-brane is purely classical, and can be determined at any point in K\"ahler moduli space.  Using open string mirror symmetry, one can hence count holomorphic disks on the A-side by mapping the problem to a classical question on the mirror B-side. For compact Calabi-Yau manifolds, this line of research was initiated in \cite{Walcher:2006rs,Morrison:2007bm} by formulating the on-shell problem ({\textit{i.e.} at the critical point of the open-string modulus) for the mirror quintic. Other one parameter models of compact Calabi-Yau hypersurfaces and complete intersections have been studied in \cite{Krefl:2008sj,Knapp:2008uw,Walcher:2009uj}. Using the technique of the variation of mixed Hodge structure to capture the open-string deformations, as first introduced in \cite{Lerche:2002ck,Lerche:2002yw}, an off-shell formulation of open string mirror symmetry for compact Calabi-Yau manifolds in the vicinity of the LG-point has been developed in \cite{Jockers:2008pe}. It has been shown in \cite{Alim:2009rf} how the notion of open string mirror symmetry extends to the geometric regime ({\textit{i.e.}} large volume phase) of the open/colsed moduli space. Moreover in \cite{Alim:2009rf}, it has been indicated how the integer open Gromov-Witten invariants of the corresponding compact A-model geometry are extracted from the non-perturbative sum of the instanton effects. In \cite{Alim:2009bx}, the integrability and flatness conditions of the Gauss-Manin connection for B-type branes have been systematically studied from a Hodge theoretic point of view. These techniques are further promoted in \cite{Jockers:2009ti} to encompass the case of heterotic string theory as well. An alternative approach to formulate the open-string deformations and to compute the effective D-brane superpotentials has been pursued in \cite{Grimm:2008dq,Grimm:2009ef}.

In this paper, we study superpotentials on D2-branes from two points of view: On the one hand, we employ techniques from \cite{Jockers:2008pe,Jockers:2009mn} to determine superpotentials in the geometric picture. Essentially, this means that we compute relative periods for B-branes by means of direct integration. One can alternatively compute the relative periods of an open/closed geometry by first deriving a set of differential equations along the lines of \cite{Aganagic:2009jq} and then solving the system of differential equations. On the other hand, we consider the Landau-Ginzburg point of the Calabi-Yau compactification and determine the superpotential using worldsheet methods. This means that we interpret the superpotential
as a generating functional for certain worldsheet correlation functions.

Since the superpotential on a B-brane is independent of the K\"ahler moduli, the results obtained in different regimes must necessarily agree. In this paper, we verify this for a class of D2 branes to first order in the bulk couplings.

This paper is organized as follows: In section 2 we first introduce the class of D2 branes we want to consider. We review how to calculate superpotential terms from the world-sheet point of view, using matrix factorizations at the Landau-Ginzburg points. Furthermore, we introduce the necessary geometric techniques to calculate superpotentials as chain integrals in a geometric setting. Section 3 is the heart of the paper, where we explicitly evaluate the bulk-boundary correlation function on the worldsheet, and compare with the corresponding geometric calculations, finding complete agreement. More specifically, we simplify in a general setting the bulk-boundary two-point function and show that the integral expression coincides with the chain integral which is responsible for the geometric effective superpotential. In section 4, we consider a concrete example and calculate superpotentials on various branches in open string moduli space. Section 5 is devoted to summary and conclusions. In appendix A, we address a technical issue regarding the relation between the effective D-brane supperpotentials and the choice of the bulk perturbations.

\section{Background and Setup}
\setcounter{equation}{0}
In this paper we want to calculate the superpotentials on branes wrapping internal 2-cycles of a Calabi-Yau manifold.

Geometrically, the starting point is the Fermat point of a hypersurface, $Y$, in a weighted projective space. We consider B-type D-branes wrapping holomorphic spheres $\IP^1 \subset Y$. At the Fermat point one can find families of such spheres, whereas at generic points in complex structure moduli space one expects to find only finitely many \footnote{In fact, the number of such holomorphically embedded curves has been counted using mirror symmetry.}. Physically, the existence of families means that a D-brane wrapping one of the curves has an open string modulus -- in other words, there exists a moduli space of supersymmetric vacua parameterized by the expectation value of this massless field. Varying the complex structure of the Calabi-Yau manifold corresponds to turning on a bulk field. The fact that a generic curve will no longer be holomorphic in the new complex structure means that this branch of the open string moduli space has been lifted by the bulk deformation. There is a non-trivial superpotential, with a diskrete set of minima, corresponding to the finitely many curves that survive the complex structure deformation.

An example for this has been studied in \cite{Baumgartl:2007an}. Here, the geometric model is the quintic 3-fold, which at the Fermat point is given by the following hypersurface in $\IP^4$
\begin{equation}
x_1^5 + x_2^5 + x_3^5 + x_4^5 + x_5^5 =0\ .
\end{equation}
A family of curves is then given by
\begin{equation}\label{eq:quinticcurves}
(x_1:x_2:x_3:x_4:x_5) = (u:\eta u:\alpha v:\beta v: \gamma v) \quad ,\quad\alpha^5+\beta^5+\gamma^5=0\quad,\quad \eta^5=-1 \ .
\end{equation}
Here, $(u:v)$ are coordinates of a $\IP^1$, and the three complex parameters $(\alpha,\beta,\gamma)$ are subject to projective equivalence and hence parameterize a one-parameter family of curves. Adding a generic degree 5 polynomial to the quintic equation, the families (\ref{eq:quinticcurves}) will no longer solve the perturbed hypersurface equation. Of particular interest are perturbations of the form
\begin{equation}
x_1^5 + x_2^5 + x_3^5 + x_4^5 + x_5^5 +x_1^3s^{(2)}(x_3,x_4,x_5) =0\ ,
\end{equation}
where $s^{(2)}$ is a degree $2$ polynomial in $(x_3,x_4,x_5)$. In this case, curves fulfilling
\begin{equation}\label{eq:quinticvacua}
\alpha^5+\beta^5+\gamma^5=0 \qquad {\rm and}  \qquad s^{(2)}(\alpha,\beta,\gamma) =0\ ,
\end{equation}
will survive the deformation. Physically, this means that there is a superpotential depending on the deformation parameters from the open and closed string sector whose minima fulfill (\ref{eq:quinticvacua})

There are various ways to determine this superpotential. First of all, in a geometrical Calabi-Yau compactification the superpotential on a brane wrapping a curve $C$ is given by an integral of the holomorphic 3-form $\Omega$ over a 3-chain $\Gamma$
\begin{equation}\label{sup0}
{\cal W}(C) - {\cal W}(C_0) = \int_\Gamma \Omega\ ,
\end{equation}
where $\partial \Gamma = C- C_0$ and $C_0$ is a reference curve. In the unperturbed situation, $C$ and $C_0$ are two members of a family of holomorphic curves. As a consequence, the above chain integral  of a $(3,0)$ form vanishes. As soon as one turns on a complex structure deformation such that the curves become non-holomorphic, the above chain integral yields a non-zero result. It will obviously depend on bulk as well as boundary parameters. The non-holomorphicity of $C$ makes it a challenge to actually evaluate the chain integral.
However, as we will review below, appropriate techniques have been developed in \cite{Jockers:2008pe}.

A second, a priori completely different approach is provided if one takes a worldsheet point of view. From this point of view, D-branes on Calabi-Yau manifolds are described by formulating boundary conditions for the underlying $N=(2,2)$ superconformal field theory. Complex structure deformations correspond to elements of the $(c,c)$ ring and D-brane moduli correspond to chiral fields of appropriate weight living at the boundary. In this context, one would regard the superpotential as a generating functional for all disk correlation functions
\begin{equation}\label{eq:Bdef}
B_{i_0\dots i_n,j_0 \dots j_m} = \langle \Phi_{j_0} \Psi_{i_0}  \int \Psi^{(1)}_{i_1} \dots \int \Psi^{(1)}_{i_n} \int \Phi^{(2)}_{j_1} \dots \int \Phi^{(2)}_{j_m} \rangle \ ,
\end{equation}
where the boundary condition is encoded by the D-brane.
Here, $\Phi_{j}$ denote  bulk fields from the $(c,c)$ ring and $\Phi_j^{(2)}$ denotes the two form descendant that can be integrated naturally over the worldsheet
\be
\Phi^{(2)}_j = [ G_{-1/2}, [ \tilde{G}_{-1/2} \Phi_j ] ] \ ,
\ee
where $G_{-1/2}$ and $\tilde{G}_{-1/2}$ are the left and right supercurrents of the $N=(2,2)$ superconformal algebra. Similarly, $\Psi_i$ denotes a chiral operator from the  open string sector whose
one form descendant can be integrated over the boundary.
\be
\Psi_i^{(1)} = [ G_{-1/2}, \Psi_i ] \ .
\ee
The integrability of disk correlators was investigated in \cite{Herbst:2004jp}. It is shown that the correlators deformed by bulk operators integrate to disk amplitudes ${\cal F}$ depending on the deformation parameters $t_j$ corresponding to the fields $\Phi_j$:
\be
B_{i_0\dots i_n,j_0 \dots j_m} = \partial_{j_1} \dots \partial_{j_m} {\cal F}_{i_0, \dots , i_n}(t) \vert_{t=0}\ .
\ee
Disk amplitudes are only cyclically symmetric in the boundary insertions, and therefore are in general not expected to be integrable with respect to the boundary insertion parameters. However, by introducing formal variables $s_i$ one can nonetheless write down a generating functional
\be
{\cal W} = \sum \frac{1}{n} s_{i_n} \dots s_{i_1} {\cal F}_{i_0, \dots , i_n}(t) \ .
\ee
The correlation functions are known to be constrained by BRST symmetry leading to interesting algebraic structures including in particular the $A_\infty$ structure obeyed by correlators without bulk insertions \cite{Herbst:2004jp,Aspinwall:2004bs}, see \cite{Ashok:2004xq,Lazaroiu:2006qc,Knapp:2008tv,Carqueville:2009ay,
Carqueville:2009xu,Aspinwall:2010mw} for recent papers exploiting algebraic structures to calculate superpotentials.

Despite the ordering ambiguities, one can still show that the following integrability condition holds \cite{Baumgartl:2007an}
\be\label{integrability}
\partial_{i_1}  B_{\Phi \Psi_{i_2} } = \partial_{i_2} B_{\Phi \Psi_{i_1}}\ .
\ee
This is a consequence of the fact that the boundary correlators are still cyclically symmetric in the boundary insertions and totally symmetric in the bulk insertions; the proof of (\ref{integrability}) then proceeds much as in the bulk case \cite{Dijkgraaf:1990dj}.
As a consequence, one expects that the bulk-boundary correlator can be integrated
in order to obtain a generating function.

In this paper, we consider examples where initially there are one-dimensional branches of the open string moduli space. Quite generally in the class of models we will consider, there will be different branches in moduli space, intersecting at special points. To give a simple example, consider again the quintic. Obviously, there are other branches similar to (\ref{eq:quinticcurves}) which are obtained by simply permuting the variables, e.g. as in
\be
(x_1:x_2:x_3:x_4:x_5) = (\alpha' v: \beta' v: u: \eta u: \gamma' v) .
\ee
This family shares a  point with (\ref{eq:quinticcurves}). As a consequence, at the intersection point one expects to find two different massless open string fields that infinitesimally generate the translation along either family. Indeed, as it turns out, the two generators exist anywhere in moduli space, not just at the intersection point. However, one of them is obstructed to first order, exhibiting a non-vanishing 3-point function at a generic point. Even more generally,  one often finds various  fields that are massless to $0^{th}$ order at least at special points in moduli space.  We refer to \cite{Baumgartl:2007an,Baumgartl:2008qp} for a complete discussion in 1-parameter models. Naturally, all of these fields can enter in correlation functions and thereby contribute to the superpotential. In the current paper, we will always isolate one particular branch in moduli space, and study the boundary field $\Psi$ that generates the translation along the branch.
It is unobstructed at the Fermat point of the bulk geometry. Turning on a bulk perturbation $\Phi$ the branch gets lifted by a superpotential that we will calculate on the  worldsheet  by computing
\be
B_{\Phi \Psi} = \langle \Phi \Psi \rangle\ ,
\ee
at arbitrary points in the open string moduli space. By integration, this yields the superpotential responsible for lifting this branch of the moduli space to first order in the bulk perturbation. Since the result is calculated at any point in boundary moduli space, it is valid to all orders in the boundary couplings.

By general arguments \cite{Brunner:1999jq} the superpotential on B-type branes is independent of the K\"ahler moduli of a Calabi-Yau compactification. We have sketched above how it can be obtained in the geometric regime as a generalized period integral and in the stringy regime as a generating functional for worldsheet correlation functions. Since the K\"ahler moduli decouple, the two expressions must be the same. We will show in section 3 that this is indeed the case \footnote{In this context, we would like to refer to the earlier work \cite{Douglas:2002fr}. In that paper, certain superpotential corrections calculated at the Landau-Ginzburg point were given a geometric interpretation.}. In the remaining part of this section, we will discuss the geometric and worldsheet techniques that we are going to employ in somewhat more detail.

\subsection{The Landau-Ginzburg Approach}

In the stringy regime of the K\"ahler moduli space of a Calabi-Yau manifold the compactification can be described by a Landau-Ginzburg orbifold. This is a two-dimensional $N=(2,2)$ superconformal field theory, specified by its F-term or superpotential $W$. The field content consists of chiral superfields with possibly different charges under the $U(1)$ R-symmetry. Relating the LG model to the geometric model, one simply replaces each coordinate by a chiral superfield, and the equation $W=0$ agrees with the hypersurface equation of the Calabi-Yau compactification. Concretely, and as is well known, the quintic at its Fermat point corresponds to a Landau-Ginzburg model with superpotential
\be
W= x_1^5 +x_{2}^{5}+x_{3}^{5}+x_{4}^{5}+x_5^5 \ ,
\ee
where all the fields have charge
$q=2/5$. The orbifold group is $\Z_5$ and projects on integer charges in the internal part of the compactification.
Another model that will play a role in this paper is the model that
geometrically is the hypersurface $\IP^{4}_{(1,1,2,2,2)}[8]$, with superpotential
\be
W= x_1^8+x_2^8 + x_3^4 + x_4^4 + x_5^4\ ,
\ee
where the charges are
$q_1=q_2=1/4$ and $q_3=q_4=q_5=1/2$; the orbifold is $\Z_8$ in this case.
B-type branes in Landau-Ginzburg models are described by matrix factorizations of the superpotential \cite{Warner:1995ay,Kapustin:2002bi,Brunner:2003dc,Lazaroiu:2003zi}. Such factorizations are specified by boundary BRST operators that square to the superpotential
\be
Q^2 = W {\bf 1} \ .
\ee
For Landau-Ginzburg orbifolds, invariant factorizations obtain an additional representation label.
A large class of examples for such BRST operators can be found by decomposing the superpotential as
\be\label{eq:ourMF}
W= \sum_i J_i E_i = \sum_i W_i, \quad {\rm where} \ W_i=J_i E_i \ .
\ee
Here, $J_i$ and $E_i$ are polynomials, and in our examples $i$ takes the values $\{0,1,2\}$. $Q$ can then be expressed as a linear combination of boundary fermions
\be
Q= \sum_i (\pi^i J_i + \bar\pi^i E_i)\ ,
\ee
where the fermions form a representation of the Clifford algebra
\be
\{ \pi^i, \bar\pi^j \} = \delta^{ij} , \quad \{\pi^i, \pi^j \} = \{ \bar\pi^i, \bar\pi^j \} =0 \ .
\ee
Using these commutation relations, one easily verifies that $Q$ squares to $W$.
The open string spectrum is now given by the cohomology of $Q$, and in particular is $\Z_2$ graded, consisting of boundary bosons and fermions. Taking into account the orbifold action, the open string spectrum consists of open string states with integer charges, whereas between branes carrying different representation labels the fractional part of the open string charges is fixed and depends only on the difference between the representation labels. In this paper, we restrict to single branes, and hence the only effect of the orbifold projection in the open string sector is to restrict to states with integer charge.

To determine the charge, note that any open string state is a linear combination of monomials in the coordinate fields $x_i$ and boundary fermions $\pi_i, \bar\pi_i$. The charges of the fermions can be read off since $Q$ squares to $W$, and $W$ is conventionally assigned $R$-charge $2$.

To calculate correlation functions between bulk and boundary fields, one can now make use of the Kapustin-Li formula \cite{Kapustin:2003ga,Herbst:2004ax}, which for a single bulk insertion $\Phi$ and a single boundary insertion $\Psi$ takes the form
\be \label{eq:KL}
\langle \Phi \Psi \rangle =  \frac{1}{n!} \oint \frac{\Phi{\rm STr}[(dQ)^n \Psi]}{\partial_1 W \dots \partial_n W} \ .
\ee
Here, $n$ is the number of variables, in our examples $n=5$ and the integrand is a meromorphic $n$-form in flat space.
The integral is a residue integral where we have dropped all factors of $2\pi i$, and in particular the contour encircles all the critical points of $W$.
The Kapustin-Li formula directly provides a method to calculate a first order term (in both bulk and boundary couplings) of the superpotential.



In the current paper, we will make use of the Kapustin-Li formula in a situation where we have families of D-branes, described by families of matrix factorizations. Consider a one-dimensional moduli space, so that the boundary BRST operator $Q$ depends on a single complex parameter $\beta$. There is then a fermion $\Psi_{\beta}$ which infinitesimally generates this deformation
\eqn{
	\Psi_\beta &\equiv \frac{\partial}{\partial\beta} Q(\beta)\ .
}

The Kapustin-Li formula then allows to obtain a result to all orders in the boundary coupling. This is achieved, since $\Psi_\beta$ generates the deformation at any point in moduli space. Hence, if one inserts $\Psi_\beta$ in (\ref{eq:KL}) one obtains a family of correlation functions \footnote{Provided the bulk deformation has been chosen such that the correlator does not vanish} and correlators with arbitrary number of insertions of $\Psi_\beta$ can be obtained by taking derivatives. Hence all coefficients $B_{i_0, \dots, i_n, j_0}$ of (\ref{eq:Bdef}) can be calculated this way \footnote{Note that our analysis is on the level of the topological theory, see \cite{gaberdiel:2009hk} for a discussion of bulk perturbations in the presence of boundaries on the level of the full conformal field theory}.

To compare results for the superpotentials in the stringy and geometric regime, one first needs to know which matrix factorization corresponds to which geometric brane. This question has been answered in \cite{Herbst:2008jq}. The basic idea is that the Landau-Ginzburg and non-linear-sigma model are different points in the moduli space of a two-dimensional gauge theory, the gauged linear sigma model. Branes of the linear sigma model can then be interpreted at the two different limit points. Cutting a long story short, starting with a matrix factorization of the Landau-Ginzburg model, one needs a lift to the underlying linear sigma model, where the discrete orbifold group is promoted to the gauge group of the linear sigma model. Interpreting this brane at large volume, one obtains a geometric D-brane. In the case at hand, these geometric branes are given by the zero set of
three polynomials $J_i$
\be
J_0=J_1=J_2=0 \ .
\ee
For the example of the quintic, the necessary calculation that shows this was explicitly performed in \cite{Herbst:2008jq}, and in the case of other examples the result follows by analogous considerations. It is important to note that the boundary BRST operator $Q$ (squaring to $W$) of the Landau-Ginzburg regime becomes a differential operator at large volume squaring to zero. The reason is  that the superpotential of the LG theory requires at large volume that the theory is restricted to a hypersurface with equation $W=0$. The open strings, which were described in terms of the BRST cohomology of the twisted differential operator $Q$ at the Landau-Ginzburg point, are now given by the Ext groups of the geometric complex.

\subsection{The Geometric Approach}

As is well-known, type II string compactifications on Calabi-Yau manifolds, in the presence of space-filling D-branes, result in
effective ${\cal N}=1$ supergravity theories in the low energy regime. These space-filling D-branes are extended through the four non-compact directions of space-time, and along the internal compactification space, they wrap even dimensional cycles of the Calabi-Yau manifold. The resulting effective superpotential of ${\cal N}=1$ supergravity theory arises from the dimensional reduction of the holomorphic Chern-Simons action to the world-volume of the D-brane along the internal space \cite{Witten:1992fb,Witten:1997ep,Aganagic:2000gs}. The holomorphic action of the Chern-Simons theory depends on the D-brane open-string moduli for the deformations of the embedding wrapped cycle, as well as the complex structure moduli of the ambient Calabi-Yau space through its coupling to the holomorphic three-form. Therefore, as a result, the effective superpotential of the D-brane depends on both closed- and open-string moduli of the D-brane configuration.
\par
In the present discussion, we are interested in space-filling D2-branes which wrap two-cycles of the internal Calabi-Yau space. The induced classical effective superpotentials of such D-brane configurations can be described in terms of certain chain integrals. These chain integrals are defined by
\begin{eqnarray}\label{relper}
\hat{\Pi}^{a}(\psi,\xi)=\int_{\hat{\Gamma}^{a}(\xi)}\Omega(\psi)\ ,
\end{eqnarray}
where $\Omega$ is the holomorphic three-form of the Calabi-Yau $Y$, and $\psi$ and $\xi$ are the closed- and open-string moduli respectively. In (\ref{relper}), $\{\hat{\Gamma}^{a}\,|\,a=1,\cdots,dim( H_{3}(Y,C,{\mathbb{Z}}))\}$ forms a basis for the three-chains that have nontrivial boundaries, lying in the two-cycle $C$ which is wrapped by the D2-brane. The classical D2-brane superpotential (\ref{sup0}) is then expressed as a linear combination of $\hat{\Pi}^{a}$ as was discussed in \cite{Aganagic:2000gs,Aganagic:2001nx,Mayr:2001xk}
\begin{eqnarray}\label{suplincom}
{\cal W}(\psi,\xi)=\hat{N}_{a}\hat{\Pi}^{a}(\psi,\xi)\ .
\end{eqnarray}
Here, $\hat{N}_{a}$ specifies the homology class of the two-cycle $C$, wrapped by the D2-brane in the ambient Calabi-Yau internal space. For the case of local Calabi-Yau manifolds, it has been shown in \cite{Aganagic:2001nx} that once one figures out the cycle which gives rise to the effective superpotential, one can construct the correct two-cycle which corresponds to the open-string mirror map. We also notice that although (\ref{suplincom}) is a  purely classical superpotential for a given D-brane configuration in the B-model, it predicts a highly nontrivial sum of non-perturbative instanton effects in the mirror A-model \cite{Ooguri:1999bv,Kachru:2000an}.
\par
As we mentioned before, space-filling D2-branes generically wrap non-holomorphic two-cycles of the Calabi-Yau internal space. In fact, holomorphic two-cycles of the Calabi-Yau $Y$ are in a one to one correspondence with the critical points of the effective D2-brane superpotential. This implies that once we consider deformations around the isolated critical points (vacua of the ${\cal N}=1$ theory) of the superpotential, we are dealing with non-holomorphic cycles. However, in order to avoid the difficulties arising from the non-holomorphicity of the two-cycles under consideration, one can employ the arguments of \cite{Lerche:2002ck,Lerche:2002yw} and replace the non-holomorphic two-cycle $C$ by a holomorphic divisor $D$ of the ambient Calabi-Yau space $Y$. The requirement for the holomorphic divisor $D$ is that it must encompass the two-cycle $C$. Upon this replacement, one is able to formulate the variational problem of handling the moduli dependence of the superpotential in terms of purely holomorphic constituents.
\par
One convenient way of setting up the variational problem is to employ the relative (co)homology groups. For the divisor $D$, embedded in the Calabi-Yau $Y$, $i:D\hookrightarrow Y$, the space of relative forms $\Omega^{*}(Y,D)$ is defined via the following short exact sequence
\begin{eqnarray}\label{shortseq}
0\longrightarrow\Omega^{*}(Y,D)\hookrightarrow\Omega^{*}(Y){\buildrel i^* \over \longrightarrow}\,\Omega^{*}(D)\longrightarrow0\ .
\end{eqnarray}
The relative cohomology groups are then defined as the space of closed modulo exact relative forms with respect to the de Rham differential operator. From the associated long exact sequence of (\ref{shortseq}) on the level of cohomology, one finds the following decomposition for the relative middle cohomology group
\begin{eqnarray}\label{decomp}
H^{3}(Y,D)\simeq\mbox{ker}\big(H^{3}(Y)\rightarrow H^{3}(D)\big)\oplus\mbox{coker}\big(H^{2}(Y)\rightarrow H^{2}(D)\big)\ .
\end{eqnarray}
The above decomposition, (\ref{decomp}), allows one to represent a relative three-form, $\underline{\Theta}$, as a pair of a closed three-form $\Theta\in H^{3}(Y)$ and a closed two-form $\theta\in H^{2}(D)$
\begin{eqnarray}\label{relform}
H^{3}(Y,D)\ni\,\underline{\Theta}=(\Theta,\theta)\ .
\end{eqnarray}
Since the above relative three-forms are elements of cohomology, they are subject to the following equivalence relation
\begin{eqnarray}\label{equiv}
\underline{\Theta}\sim\underline{\Theta}+(d\mu,i^{*}\mu-d\nu)\ ,
\end{eqnarray}
where $\mu$ is a two-form on the Calabi-Yau $Y$, and $\nu$ is a one-form defined on the divisor $D$. Imposing the above equivalence relation on closed relative three-forms, one can pair relative three-forms and three-chains in a well-defined manner. As a consequence, we can define the notion of relative periods as the integral of the relative holomorphic three-form of the Calabi-Yau over a relative homology basis $\underline{\Gamma}^{a}$ as
\begin{eqnarray}\label{relperiod}
\underline{\Pi}^{a}(\psi,\xi)=\int_{\underline{\Gamma}^{a}}\underline{\Omega}(\psi,\xi)
\qquad,\qquad \underline{\Gamma}^{a}\in H_{3}(Y,D,{\mathbb{Z}})\ .
\end{eqnarray}
In the above formula, the moduli dependence of the relative periods is entirely captured by the relative three-form $\underline{\Omega}$. The effective D2-brane superpotential is now a linear combination of the relative periods. Therefore, one can trace the moduli dependence of the superpotential through the dependence of the relative periods on closed- and open-string moduli.
\par
The appropriate formalism to trace the moduli dependence of relative periods is the variation of mixed Hodge structure \cite{Lerche:2002ck,Lerche:2002yw,Jockers:2008pe}. Cutting a long story short, in order to define the variation of mixed Hodge structure, one needs two ingredients \footnote{\ For a rigorous mathematical definition of the variation of mixed Hodge structure, we refer the reader to \cite{Voisin}.}. The first ingredient is a finite decreasing weight filtration $F^{p}$ on the complexified relative middle cohomology group $H^{3}(Y,D,{\mathbb{C}})$
\begin{eqnarray}\label{filt}
F^{p}=F^{p+1}\oplus H^{p,3-p}(Y,D)\quad,\quad\mbox{where \ \ } p=3,2,1,0\, ,\qquad\mbox{and}\qquad F^{4}=0\ .
\end{eqnarray}
The second ingredient to define the variational problem is a finite weight increasing filtration $W_{p}$ on the rational relative middle cohomology group.  This weight filtration is induced from the decomposition (\ref{decomp}) and it reads
\begin{eqnarray}\label{filtW}
W_{2}=0\quad,\quad W_{3}=H^{3}(Y)\quad,\quad  W_{4}=H^{3}(Y,D)\quad,\quad  W_{4/3}=\frac{W_{4}}{W_{3}}\simeq H^{2}(D)\ .
\end{eqnarray}
Since the whole moduli dependence of the relative periods is captured by the relative forms, to address the moduli dependence of the relative periods, we can analyze the behavior of the relative three-forms under infinitesimal deformations. Due to Griffiths transversality, the infinitesimal variation of complex structure (closed-string) moduli, $\partial_{\psi}$, changes the Hodge-type of the form. On the other hand, the infinitesimal variation of the open-string moduli, $\partial_{\xi}$, only affects the two-form part of the relative form. After all, the systematic action of the infinitesimal variation of open- and closed-string moduli on the defined mixed Hodge structure is summarized in the following diagram
\begin{equation}\label{VHS}
\xymatrix{
  F^3\cap W_3 \ar[r]^{\partial_\psi} \ar[rd]^{\partial_\xi}& F^2\cap W_3  \ar[r]^{\partial_\psi}  \ar[rd]^{\partial_\xi}
     & F^1\cap W_3  \ar[r]^{\partial_\psi} \ar[rd]^{\partial_\xi} & F^0\cap W_3 \ar[d]_{\partial_{\xi}} \nonumber\cr
  & F^2\cap W_{4/3} \ar[r]^{\partial_\psi,\partial_\xi} &  F^1\cap W_{4/3} \ar[r]^{\partial_\psi,\partial_\xi} & F^0 \cap W_{4/3} \ .}
\end{equation}
As the above diagram exhibits, if one applies the infinitesimal variational operators $\partial_{\psi}$, and $\partial_{\xi}$ successively, one finds that after a finite number of steps, one saturates the relative cohomology group, and in this way, one generates a set of differential operators which annihilate the relative periods. Hence, solving the set of differential equations simultaneously, we find the relative periods as functions of both open- and closed-string moduli. For further information on the structure of the system of differential equations, and other related issues of the $N=1$ special geometry structure, we refer the reader to \cite{Lerche:2002ck,Lerche:2002yw,Lerche:2001cw} as well as \cite{Jockers:2008pe,Alim:2009rf,Alim:2009bx}. In particular, it is shown in \cite{Alim:2009rf} how one can economically generate the system of linear differential equations, using the techniques of toric geometry.

\section{D-brane Superpotentials}
\setcounter{equation}{0}
As mentioned in the previous section, D-brane superpotentials arising from type II string compactifications on Calabi-Yau manifolds can be computed in two different ways. On the one hand, from the worldsheet point of view, the superpotential is regarded as the generating functional of all disk correlation functions with appropriate bulk and boundary insertions. On the other hand, from the geometric perspective, the effective D2-brane superpotential is captured by the relative periods of the corresponding open/closed geometry. In the first part of this section, we start by simplifying the integral expression of the bulk-boundary disk correlator given by the Kapustin-Li formula. In the second part of this section, we simplify the  geometric chain integral expression of the effective superpotential. Comparing the two simplified integral expressions from the worldsheet and geometric perspectives, we prove that they reduce to the same expression, and therefore they coincide precisely.

\subsection{Superpotentials from Kapustin-Li Formula}

Using the Kapustin-Li formula, in this section we derive a convenient expression for the bulk-boundary correlation function $\langle G\Psi_\beta\rangle$ which gives rise to the effective superpotential of a given D2-brane configuration. In this derivation, we start by evaluating (\ref{eq:KL}) for matrix factorizations of the type (\ref{eq:ourMF}), where the boundary deformation is identified with the derivative of the boundary BRST operator, $\Psi_\beta=\partial_\beta Q(\beta)$, and $G$ represents the chosen bulk deformation whose role is to lift the open string moduli space. In the class of models under consideration, the criterion for choosing the deformation $G$ is that it does not excite other obstructed fermions. The differential operator $d$ appearing in (\ref{eq:KL}) acts on $Q$ by
\eqn{
	dQ &= \sum_{i=0,1,2}(\pi_i dJ_i + \bar\pi_i dE_i)\ .
}
We notice that $\pi_i dJ_i$ and $\bar\pi_{i} dE_i$ are bosonic. We can now write the Kapustin-Li formula as
\eqn{\label{buboco}
	\langle G \Psi\rangle
	&= \frac{1}{5!}\oint_{{\cal K}_W} \frac{G }{ \partial_{x_1} W  \partial_{x_2} W  \partial_{x_3} W  \partial_{x_4} W  \partial_{x_5} W} \int d^{6}\pi \Psi\, d Q^{5} \ ,
}
The integral is performed over closed contours ${\cal K}_W$ which encircle the critical points of the LG superpotential $W$. Therefore, the bulk-boundary correlation function is computed by taking the residues of the integrand of (\ref{buboco}) around the isolated critical points of $W$. In order to compute expressions of the type (\ref{buboco}) which involve $\Psi_\beta$, we rewrite $d\beta(\partial_\beta Q) (d Q)^{5}$ in a more convenient form in which the modulus $\beta$ is treated as an additional coordinate. We introduce a new differential $\hat{d}$ which is given by $\hat d = d + d\beta\partial_\beta$. Thus, using the new differential form on this extended space, we have
\eqn{
	\frac{1}{6!}(\hd Q)^{6} = \pi_0\hd J_0 \pi_1\hd J_1\pi_2\hd J_2 \bar\pi_0\hd E_0\bar\pi_1\hd E_1\bar\pi_2\hd E_2\ .
}
After integrating out the boundary fermion or, equivalently, by taking the supertrace \cite{Takayanagi:2000rz}, we gain
\eqn{
\label{5trace}
	\frac{1}{6!}\int d^{6}\pi (\hd Q)^{6}
	= \hd J_0 \wedge \hd J_1 \wedge \hd J_2 \wedge \hd E_0 \wedge \hd E_1 \wedge \hd E_2\ .
}
Using (\ref{eq:ourMF}), we can now reexpress the differentials $\hd E_i$ in terms of $J_i$ and $W_i$ in the following way:
\eqn{
	\hd E_i = \frac{\hd W_i}{J_i} - \frac{E_i\hd J_i}{J_i}\ .
}
We can now use the above formula to eliminate the $E_{i}$ in favor of $W_{i}$. This brings (\ref{5trace}) into a more convenient form
\eqn{
	\frac{1}{6!}\int d^{6}\pi (\hd Q)^{6} = \frac{\hd J_0}{J_0} \frac{\hd J_1}{J_1} \frac{\hd J_2}{J_2} \hd W_0 \hd W_1 \hd W_2\ .
}
The fact that the $E_i$'s disappear in the above expression indicates that as far as the the bulk-boundary correlation function is concerned, all information is solely contained in the $J_i$'s \footnote{One reason, why this is expected is that the branes under consideration are described by 2-cycles specified by the equations $J_i=0$, and clearly the choice of $E_i$ does not enter at large volume.}. The full expression for integrated bulk-boundary correlator is found \footnote{In fact, this formula is valid even more generally. Since other fermions in the spectrum can be related to $\partial_\beta Q$ by multiplication with a certain set of monomials involving negative exponents, $\partial_\beta Q$ may be replaced by $f(x) \partial_\beta Q$, where $f = \prod_i x_i^{a_i}$, $a_i \in {\mathbb Z}$  so that $f(x)$ has vanishing charge, in order to compute correlators with fermions other than the exactly marginal one. See \cite{Baumgartl:2008qp} for examples.}

\eqn{
\label{KL2}
	\int_{\beta_0}^\beta d\beta \, \langle G \partial_\beta Q\rangle
	&= \int_{{\cal K}_W\times {\cal I}_\beta}  \frac{G}{ \partial_{x_1} W  \partial_{x_2} W  \partial_{x_3} W  \partial_{x_4} W  \partial_{x_5} W}\frac{\hd J_0}{J_0} \frac{\hd J_1}{J_1} \frac{\hd J_2}{J_2} \hd W_0 \hd W_1 \hd W_2\ .
}
The integral on the right hand side of (\ref{KL2}) is performed over closed contours ${\cal K}_{W}$ which encompass the critical points of $W$ while the integration on the modulus $\beta$ is performed over an open interval ${\cal I}_\beta$. We would like to stress that $\hd$ acts on $J_i$ as well as $W_i$. Although the whole $W$ is independent of open modulus $\beta$, the individual pieces depend on $\beta$.

\bigskip

Now, we specialize our discussion to the case of deformations around the Fermat point. In particular, we assume that the LG superpotential is given by a quasi-homogeneous polynomial of degree $D$
\eqn{
	W &= x_1^{A_1}+x_2^{A_2}+x_3^{A_3}+x_4^{A_4}+x_5^{A_5}\ ,
}
in the weighted projective space ${\mathbb P}^{4}_{(\frac{D}{A_1},\frac{D}{A_2},\frac{D}{A_3},\frac{D}{A_4},\frac{D}{A_5})}[D]$, where the Calabi-Yau condition dictates that the degree $D$ is the sum over the weights $D/A_i$. The D2-brane wraps a ${\mathbb P}^1$ in the ambient ${\mathbb P}^4$ and is described as the common zero locus of three holomorphic polynomials $J_i$. The explicit forms of $J_i$'s are given by
\eqn{
\label{Jcons}
	J_0 &= x_1-\eta x_2^e
	\qquad,\qquad
	J_1 = \alpha^b x_4 - \beta x_3^b
	\qquad,\qquad
	J_2 = \alpha^c x_5 - \gamma x_3^c \ ,
}
where the  integers $e, b, c$ are defined such that
\eqn{
	A_3 = bA_4 = c A_5 \qquad ,\qquad A_2=e A_1\ .
}
The latter  conditions simply ensure that (\ref{Jcons}) are homogeneous equations. We get a further constraint on the parameters $(\alpha:\beta:\gamma)$ from the condition that $J_i=0$ solves the equation $W=0$. In matrix factorization language, this requirement ensures (using the Nullstellensatz) that $W$ can be factorized as $W=\sum J_i E_i$.
Geometrically, this translates into the condition  that the embedded ${\mathbb{P}}^{1}$ lies in the Calabi-Yau target space. Either way, we arrive at  the following constraint
\eqn{
\label{OSMS}
	\alpha^{A_3} + \beta^{A_4} + \gamma^{A_5} = 0
	\qquad,\qquad
	\eta^{A_1}=-1\ ,
}
where $(\alpha:\beta:\gamma)$ characterize the homogeneous coordinates of a weighted ${\mathbb P}^2$. The above equations then describe a Riemann surface whose genus depends on the weights of the coordinates. It is interpreted as the open string moduli space.

To spell out the factorization more explicitly, we split the LG superpotential $W$ into three parts $W=W_0+W_1+W_2$. Given the holomorphic constraints $J_i$, one can explicitly derive compact expressions for $E_i$ and $W_i$ such that $W_i=J_iE_i$
\eqn{
	E_0 = \sum_{m=0}^{A_1-1} \eta^m x_1^{A_1-m-1}x_2^{em} \quad,\quad
	E_1 = \sum_{m=0}^{A_4-1} \frac{\beta^m}{\alpha^{bm+b}} x_3^{bm}x_4^{A_4-m-1} \quad,\quad
	E_2 = \sum_{m=0}^{A_5-1} \frac{\gamma^m}{\alpha^{cm+c}} x_3^{cm}x_5^{C-m-1}\ ,
}
and
\eqn{\label{W012}
	W_0(x_1, x_2) &= x_1^{A_1}+x_2^{A_2}\ , \\
	W_1(x_3, x_4; \alpha,\beta) &= f(\alpha,\beta)\, x_3^{A_3} + x_4^{A_4}\ , \\
	W_2(x_3, x_5; \alpha, \gamma) &= g(\alpha, \gamma)\, x_3^{A_3} + x_5^{A_5}\ .
}
In the above expression (\ref{W012}), the moduli-dependent functions $f$ and $g$ are defined in terms of $\beta$ and $\gamma$ as
\eqn{
	f(\beta) &= -\alpha^{-A_3}\beta^{A_4} \qquad \text{and}\qquad
	g(\beta) = -\alpha^{-A_3}\gamma^{A_5}\ .
}
Of course this splitting is designed in a way such that the whole LG superpotential $W$ is independent of the open string modulus. Since $\alpha$, $\beta$ and $\gamma$ are coordinates in the homogenous space $\P^2$, we can always set one of these coordinates to one by choosing an appropriate patch. Without loss of generality, let us assume $\alpha=1$. By the constraint (\ref{OSMS}), we recognize that both $\beta$ and $\gamma$ cannot be regarded as free parameters, and one of them can be eliminated in favor of the other. Let us regard $\beta$ as the free parameter which plays the role of the open-string modulus. This implies $\gamma=\gamma(\beta)$.

In the next step, we want to simplify (\ref{KL2}) for Fermat type models.
First, we rescale the coordinate $x_3$ by a factor of $\gamma$, i.e.\
\eqn{
\label{x3til}
	x_3 \mapsto \gamma^{\frac{1}{c}}\, x_3\ .
}
Under this rescaling,  (\ref{KL2}) is invariant except for a factor of $\gamma^{-\frac{1}{c}}$ which stems from the transformation of the derivative with respect to $x_3$ in the denominator.
In order to be more explicit, we extract $d\beta$ from the above formula. After some simplifications, we have
\eqn{
\label{Geq}
	\int d\beta \langle G \Psi\rangle
	=& -\int \frac{d\beta}{\gamma^\frac{1}{c}}\oint_{{\cal K}_W}\,G\,\frac{\partial_{\beta}J_{1}}{J_{1}}\frac{dJ_{0}}{J_{0}}\frac{dJ_{2}}{J_{2}}
			 \frac{dW_{0}\,dW_{1}\,dW_{2}}{\partial_{x_{1}}W\partial_{x_{2}}W\partial_{x_{3}}
			W\partial_{x_{4}}W\partial_{x_{5}}W}\\
		&+\int \frac{d\beta}{\gamma^\frac{1}{c}}\oint_{{\cal K}_W}\,G\,\frac{\partial_{\beta}J_{2}}
			{J_{2}}\frac{dJ_{0}}{J_{0}}\frac{dJ_{1}}{J_{1}}
			 \frac{dW_{0}\,dW_{1}\,dW_{2}}{\partial_{x_{1}}W\partial_{x_{2}}W\partial_{x_{3}}
			W\partial_{x_{4}}W\partial_{x_{5}}W}\\
		&+\int \frac{d\beta}{\gamma^\frac{1}{c}}\oint_{{\cal K}_W}\,G\,\frac{dJ_{0}}{J_{0}}\frac{dJ_{1}}{J_{1}}
			\frac{dJ_{2}}{J_{2}}\frac{\partial_{\beta}W_{1}\,dW_{0}\,dW_2}
			{\partial_{x_{1}}W\partial_{x_{2}}W\partial_{x_{3}}
			W\partial_{x_{4}}W\partial_{x_{5}}W}\ .
}
The first thing we notice is that the second term in (\ref{Geq}) is zero, because once we have performed the change of coordinates (\ref{x3til}), $J_{2}(x)$ is independent of moduli (i.e. $\partial_{\beta}J_{2}=0$). Also note that via (\ref{x3til}), $W_{2}$ becomes independent of the open modulus, but in turn, the whole $W$ will depend on the open modulus $\beta$. Now, we show that based on a simple power counting argument, the third term in (\ref{Geq}) does not give any contribution upon taking the residues.

For simplicity, let us for the moment assume that the ambient space is an ordinary projective space whose coordinates all have the same weights (this can be easily generalized to the case of weighted projective spaces as well).
To see this, first note that the whole dependence in $ {x}_{1}$ and $ {x}_{2}$ in the differentials comes from $dW_{0}$ and $dJ_{0}$. Let us split $G$ in the following way $G( {x})=s_{1}( {x}_{1}, {x}_{2})\,s_{2}( {x}_{3},
 {x}_{4}, {x}_{5})$. Obviously, $dW_{0}=N {x}_{1}^{N-1}d {x}_{1}+N {x}_{2}^{N-1}d {x}_{2}$, where $N$ is the degree of the defining equation $W$ in terms of the homogenous coordinates of the ambient projective space. Let us now first take the first term $N {x}_{1}^{N-1}d {x}_{1}$. It is clear that $  x_{1}^{N-1}$ will be canceled by $\partial_{  x_{1}}W$ in the denominator. Therefore, in order to have a nonzero result, $J_{0}$ in the denominator has to provide a factor of $  x_{1}$. The only way to have a non-vanishing result upon taking residue with respect to $ {x}_{2}$ is that $s_{1}( {x}_{1}, {x}_{2})$ is a polynomial of degree $N-2$, so that only a factor of $ {x}_{2}$ remains in the denominator when we consider $\partial_{  x_{2}}W$. We arrive at the same conclusion if we consider the second term $N {x}_{2}^{N-1}d {x}_{2}$. After all, this means that $s_{2}( {x}_{3}, {x}_{4}, {x}_{5})$ has to have degree 2, in order to preserve the homogeneity property of the defining equation $W$. But if this the case, this means that there is no way to cancel both $ {x}_{4}^{N-1}$ and $ {x}_{5}^{N-1}$ coming from $\partial_{  x_{4}}W$ and $\partial_{  x_{5}}W$ factors respectively in the denominator of (\ref{Geq}). Therefore, it is clear that the third term of (\ref{Geq}) does not give any contribution, and this argument can be easily generalized for the case of weighted projective space.
%
%
Thus we get for the correlation function
\eqn{
	\label{Geq2}
	\int_{\beta_0}^\beta d\beta \, \langle G \partial_\beta Q\rangle =-\int_{\beta_0}^\beta \frac{d\beta}{\gamma(\beta)^\frac{1}{c}}\oint_{{\cal K}_W}\,G\,\frac{\partial_{\beta}J_{1}}
{J_{1}}\frac{dJ_{0}}{J_{0}}\frac{dJ_{2}}{J_{2}}
\frac{dW_{0}\,dW_{1}\,dW_{2}}{\partial_{x_{1}}W\partial_{x_{2}}W\partial_{x_{3}}
W\partial_{x_{4}}W\partial_{x_{5}}W}\ .
}

In the above expression, the modulus which appears in $J_{1}$ is $\frac{\beta}{\gamma^{b/c}}$. Therefore, it is convenient to define a new modulus $\xi\equiv\frac{\beta}{\gamma^{b/c}}$ and rewrite the above integral in terms of $\xi$. Note that $d\beta\partial_{\beta}
=d\xi\partial_{\xi}$, so the correlation function can be expressed in terms of $\xi$ as
\eqn{\label{cor1}
	\int_{\xi_0}^\xi d\xi \, \langle G \partial_\xi Q\rangle =-\int_{\xi_0}^\xi \frac{d\xi}{\gamma(\xi)^{\frac{1}{c}}}\oint_{{\cal K}_W}\,G\,
	\frac{\partial_{\xi}J_{1}}
	{J_{1}}\frac{dJ_{0}}{J_{0}}\frac{dJ_{2}}{J_{2}}
	\frac{dW_{0}\,dW_{1}\,dW_{2}}{\partial_{x_{1}}W\partial_{x_{2}}W\partial_{x_{3}}
	W\partial_{x_{4}}W\partial_{x_{5}}W}\ .
}
As a next step we make the following change of variables:
\eqn{
\label{chvar}
{x}_{1}\mapsto  {x}_{1}\quad,\quad  {x}_{2}\mapsto v=J_{0}( {x}_{1}, {x}_{2})\quad,\quad  {x}_{3}\mapsto w=J_{2}( {x}_{3}, {x}_{5})\quad,\quad  {x}_{4}\mapsto  {x}_{4}\quad,\quad  {x}_{5}\mapsto  {x}_{5}\ .
}
This implies that
\eqn{
\label{der0}
	dW_0 &= \partial_{x_{1}}W d {x}_{1}+\partial_{v}W_{0}\,dv \\
	dW_1 &= \partial_{x_{4}}W d {x}_{4}+\partial_{w}W_{1}\,dw\ ,\\
	dW_2 &= \partial_{x_{5}}W d {x}_{5}+\partial_{w}W_{2}\,dw\ .
}
Substituting (\ref{der0}) into (\ref{cor1}), we find
\eqn{
\label{com}
	\int_{\xi_0}^\xi d\xi \, \langle G \partial_\xi Q\rangle=&-\int_{\xi_0}^\xi \frac{d\xi}{\gamma(\xi)^\frac{1}{c}}\oint_{{\cal K}_W}\,G\,
	\frac{\partial_{\xi}J_{1}}{J_{1}}\frac{dv}{v}\frac{dw}{w}\frac{(\partial_{x_{1}}W d {x}_{1})
	(\partial_{x_{4}}W d {x}_{4})(\partial_{x_{5}}W d {x}_{5})}{\partial_{x_{1}}W\partial_{x_{2}}W\partial_{x_{3}}
	W\partial_{x_{4}}W\partial_{x_{5}}W} \\
	=&-\int_{\xi_0}^\xi \frac{d\xi}{\gamma(\xi)^\frac{1}{c}}\oint_{{\cal K}_W}\,G\,
		\frac{\partial_{\xi}J_{1}}{J_{1}}\frac{dv \, dw \, d {x}_{1} \, d {x}_{4}\,
		d{x}_{5}}{v\,\partial_{ {x}_{2}}v\,\partial_{v}W\cdot w\,\partial_{ {x}_{3}}w\,\partial_{w}W}\ .
}

Another way to arrive at this result is due to the observation that in (\ref{KL2}) one can replace certain derivatives in the denominator by
$\partial_{x_1} W = \partial_{x_1}W_0$, $\partial_{x_4} W = \partial_{x_4}W_1$ and $\partial_{x_5} W = \partial_{x_5}W_2$.
Using now that under the residue integral
\eqn{
	\frac{dW_0}{\partial_{x_1}W_0} = dx_1,\qquad
	\frac{dW_1}{\partial_{x_4}W_1} = dx_4,\qquad
	\frac{dW_2}{\partial_{x_5}W_2} = dx_5
}
we directly arrive at
\eqn{
	\int_{\beta_0}^\beta d\beta \, \langle G \partial_\beta Q\rangle
	&= \int_{{\cal K}_W \times {\cal I}_\beta}  \frac{1}{\gamma^{\frac{1}{c}}}\frac{G}{ \partial_{x_2} W  \partial_{x_3} W }\frac{\hd J_0}{J_0} \frac{\hd J_1}{J_1} \frac{\hd J_2}{J_2} dx_2 dx_4 dx_5\ .
}
Since $J_0$ and $J_2$ are independent of $\beta$, we can repeat the steps above and bring the correlator into the form of (\ref{com}).

In the next section, we will show that (\ref{com}) coincides with the simplified expression for the chain integral that gives rise to the effective D-brane superpotential. In this manner, we establish an explicit map between the bulk-boundary correlation function computed by the matrix factorization technique and the superpotential chain integral in the geometric picture.

\subsection{Superpotentials from Chain Integrals}

As explained in section 2, D2-branes wrap holomorphic two-cycles of the internal Calabi-Yau space. These holomorphic cycles only exist at specific points in the closed-string moduli space, and hence become obstructed upon considering generic deformations. The obstruction generates a superpotential for the 4-dimensional effective theory, and from geometry point of view, it is captured by a certain chain integral. In this section, we simplify the chain integral which leads to the superpotential and show that it agrees with the simplified Kapustin-Li formula presented in the last section.
\par
To start, let us assume that $\beta$ is the open-string modulus associated with a given D2-brane configuration. At the Fermat point in the vicinity of the LG-point in the closed-string moduli space, there exists a family of holomorphic two-cycles. The D2-brane wraps a member of this family and can freely move without spending any energy. However, once we start deforming away from the Fermat point, the family of holomorphic two-cycles becomes obstructed. As is shown in \cite{Witten:1997ep,Aganagic:2000gs}, this leads to a superpotential for the 4-dimensional physics, which is given by a chain integral coming from the dimensional reduction of the holomorphic Chern-Simons action
\begin{eqnarray}\label{sup}
{\cal{W}}(\psi,\beta)=\int_{\Gamma(\beta)}\Omega\ .
\end{eqnarray}
In the above expression, $\Omega$ is the holomorphic three-form of the Calabi-Yau, $\psi$ is the closed-string modulus, and $\Gamma(\beta)$ is a three-chain whose ends are holomorphic two-cycles. These holomorphic two-cycles correspond to two distinct vacua of the 4-dimensional physics and the superpotential (\ref{sup}) can be regarded as a domain wall tension interpolating between these two vacua. One should notice that choice of the holomorphic three-form, $\Omega$, is unique, up to an overall closed-string dependent function. In order to remove this normalization ambiguity and to make any comparison with the worldsheet correlators, one needs to reexpress the superpotential (\ref{sup}) in appropriate flat coordinates. However, since the CFT correlators that we established in section 3.1 are computed to first order in closed-string moduli, the normalization ambiguity of the holomorphic three-form does not play any role in here, and one can directly compare (\ref{sup}) against the bulk-boundary correlator. As we will show, the bulk-boundary correlator which is first order in closed-string moduli and exact in terms of the open-string modulus is sufficient to construct the correct three-chain which is responsible for the superpotential (\ref{sup}). Once one finds out the correct three-chain, one can compute the exact effective superpotential in all moduli in the geometric picture.
\par
It is shown in \cite{Lerche:2002yw,Lerche:2002ck,Jockers:2008pe,Alim:2009rf} that the  chain integrals of the type (\ref{sup}) fulfill a set of differential equations which are governed by the variation of mixed Hodge structure. The basic idea behind this procedure is that instead of working with the absolute homology/cohomology groups, one employs relative homology/cohomology groups to capture the chain integrals of the type (\ref{sup}) by means of the variation of mixed Hodge structure. In this framework, the superpotential (\ref{sup}) is rewritten as \cite{Lerche:2002ck,Lerche:2002yw,Jockers:2008pe,Alim:2009rf}
\begin{eqnarray}\label{sup1}
{\cal{W}}(\psi,\beta)=\int_{\Gamma_{0}}\underline{\Omega}\ ,
\end{eqnarray}
where $\underline{\Omega}\in H^{3}(Y,D)$ is the holomorphic three-form of the Calabi-Yau relative to a divisor $D$ which encompasses the two-cycle that D2-brane wraps on. The whole moduli dependence of (\ref{sup1}) is encoded in the relative form and $\Gamma_{0}$ is a three-chain $\Gamma_{0}\in H_{3}(Y,D)$. As is shown in \cite{Jockers:2008pe,Li:2009dz}, the relative holomorphic three-form $\underline{\Omega}$ has an explicit representation. Let us assume that we choose a patch in which $x_{1}\neq0$. Then the superpotential (\ref{sup1}) is rewritten as
\begin{eqnarray}\label{sup2}
{\cal{W}}(\psi,\beta)=\int_{\Gamma_{0}}\frac{x_{1}\,dx_{2}dx_{3}dx_{4}}{\partial_{x_{5}}P}\,
\log\Big(\frac{Q(\beta)}{Q(\beta_{0})}\Big)=\int_{\beta_{0}}^{\beta}d\beta\int_{[C]}
\frac{\partial_{\beta}Q}{Q}\frac{x_{1}\,dx_{2}dx_{3}dx_{4}}{\partial_{x_{5}}P}\ ,
\end{eqnarray}
where $\{P(\psi)=0\}$ defines the Calabi-Yau hypersurface $Y$. The locus   $D\equiv\{Q(\beta)=0\}\cap\{P(\psi)=0\}$ is a chosen divisor in the ambient Calabi-Yau space $Y$, which encompasses the two-cycle that D2-brane wraps on. The two-cycle $C$ is the base  of the three-chain $\Gamma_{0}$, which lies in the divisor $D$, and it sweeps the whole chain as we vary the open-modulus $\beta$. Since the integration in (\ref{sup2}) is performed on a closed form,  what matters for the superpotential is only the homology class of $C$ inside the chosen divisor $D$, and any other representative of $[C]$ is equally good. Before we explicitly specify $C$, let us first fix the divisor $D$. As we mentioned before, the requirement for choosing $D$ is that it should encompass the family of holomorphic curves which exists at the Fermat point. This family of curves is specified via three holomorphic constraints $\{J_{0}(x)=0\}\cap\{J_{1}(x;\beta)=0\}\cap\{J_{2}(x;\gamma(\beta))=0\}$ in the ambient projective space. Therefore, the holomorphic constraint $Q(\beta)=0$ which defines the divisor in the current context must include at least one of the $J$-factors. In order to define a well-defined variational problem, $Q(\beta)$ must include $J_{1}$ or $J_{2}$ in order to be able to follow the members of the given family of holomorphic curves. The effective superpotential is, of course, independent of the way of embedding the family of holomorphic curves into a family of divisors. Without loss of generality, let us assume that $Q$ contains $J_{1}$ and is defined in the following way
\begin{eqnarray}\label{QD}
Q(x;\beta)=g(x)\big(J_{1}(x;\beta)\big)^{m}\ ,
\end{eqnarray}
where $g(x)$ is an arbitrary smooth function and $m$ is an arbitrary positive real number. Substituting (\ref{QD}) into (\ref{sup2}), we find for the superpotential
\begin{eqnarray}\label{sup3}
{\cal{W}}(\psi,\beta)=\int_{\beta_{0}}^{\beta}d\beta\int_{[C]}
\frac{\partial_{\beta}J_{1}}{J_{1}}\frac{x_{1}\,dx_{2}dx_{3}dx_{4}}{\partial_{x_{5}}P}\ .
\end{eqnarray}
In fact, we also get a constant prefactor $m$ in (\ref{sup3}), but we easily drop the overall numerical coefficients. The last piece we need to specify in (\ref{sup3}) is the two-cycle $C$. To define it, we first notice that the three-chain $\Gamma$ is sliced by the divisor $D$ and the two-cycle $C$ is realized as the intersection of the three-chain and the divisor. Therefore, the two-cycle $C$ sweeps the whole chain, as we vary the open modulus. Moreover, as we mentioned before, the family of holomorphic curves which defines the D2-brane geometry only exists at the Fermat point, and as soon as we perturb the Fermat polynomial by adding a deformation, the family of two-cycles ceases to be holomorphic. Therefore, $C$ has to be a family of {\textit{non-holomorphic}} two-cycles. We define $C$ in the vicinity of the LG-point in the open/closed moduli space to be
\begin{align}\label{Ccyc}
C^{(1)}=\Big\{x_{k}\,\Big|\,& x_{1}=\mbox{const.}\,\, ,\,\, \big|J_{0}(x)\big|=\big|\psi\, f_{1}(\psi)\,x_{1}\big|\,\, ,\,\, \big|J_{2}(x;\gamma(\beta))\big|=\big|\psi\, f_{2}(\psi)\big|\,\, ,\nonumber\\
&\,x_{4}\mbox{ given by the solution to }
J_{1}(x;\beta)=0\mbox{ as $\beta$ tends to zero}\ ,\\
&\,x_{5}\mbox{ given by the solution to }P(x;\psi)=0\mbox{ as $\psi$ tends to zero}\Big\}\ ,\nonumber
\end{align}
where the superscript $(1)$ indicates that the cycle $C$ has been realized in the patch $x_{1}\neq0$. In the above expression, $f_{1}(\psi)$ and $f_{2}(\psi)$ are two nowhere vanishing polynomials in the {\textit{vicinity of the LG-point}}. One can simply choose $f_{1}$ and $f_{2}$ to be constants. There are few comments in order. First, one clearly realizes that for a generic value of the closed-string modulus, $\psi$, $C$ defines a non-holomorphic two-cycle which encircles $J_{0}=0$ and $J_{2}=0$ with the radii $|\psi f_{1}\,x_{1}|$ and $|\psi f_{2}|$ respectively. Second, the fact that one coordinate of the ambient projective space is given by the solution of $P=0$ ensures that $C$ lies in the Calabi-Yau space $Y$, and similarly, the solution of $J_{1}=0$ ensures that $C$ lies in  the chosen divisor $D$. Third, one recognizes that at the Fermat point ($\psi=0$), the two circles in (\ref{Ccyc}) shrink to zero size and the holomorphic defining equations $J_{0}=J_{2}=0$ are restored. Hence, the three-chain $\Gamma_{0}$ ends on a member of the family of holomorphic two-cycles defined at the Fermat point via $J_{0}(x)=J_{1}(x;\beta)=J_{2}(x;\gamma(\beta))=0$. Fourth, one realizes that if one changes the chosen patch $(1)$, the first circle shrinks to zero size at $x_{1}=0$. This implies that the cycle $C$ has the topology of an sphere.
\par
Now, our goal is to simplify the superpotential chain integral (\ref{sup3}) and to show that it reduces to the simplified Kapustin-Li formula presented in the previous section. As the first step, we perform the same change of coordinates (\ref{x3til}) that we imposed on the Kapustin-Li formula. Upon doing this, $J_{2}$ which appears in the definition of $C$ loses its independence to moduli and becomes rigid. In turn, the modulus of $J_{1}$, defined in (\ref{Jcons}), changes to $\frac{\beta}{\gamma^{b/c}}$. It is very convenient to define a new open modulus $\xi=\frac{\beta}{\gamma^{b/c}}$ (as in the Kapustin-Li formula), and rewrite the superpotential in terms of this new modulus
\begin{eqnarray}\label{supp}
{\cal{W}}(\psi,\xi)=\int_{\xi_{0}}^{\xi}\frac{d\xi}{\gamma(\xi)^\frac{1}{c}}\int_{C}
\frac{\partial_{\xi}J_{1}( {x};\xi)}{J_{1}}\frac{ {x}_{1}d {x}_{2}d {x}_{3}
d {x}_{4}}{\partial_{ {x}_{5}}P}\ .
\end{eqnarray}
We can now massage the above expression (\ref{supp}) and bring it into a form which is not sensitive to the choice of the patch in the ambient projective space. To achieve this, we notice that
\begin{align}\label{sup4}
{\cal{W}}(\psi,\xi)=&\int_{\xi_{0}}^{\xi}\frac{d\xi}{\gamma(\xi)^\frac{1}{c}}
\oint_{ {\tilde{\gamma}_{2}}\times {\tilde{\gamma}}_{3}\times\gamma_{4}}
\frac{\partial_{\xi}J_{1}}{J_{1}}
\frac{ {x}_{1}\,d {x}_{2}d {x}_{3}d {x}_{4}}{\partial_{ {x}_{5}}P}
\nonumber\\
=&\int_{\xi_{0}}^{\xi}\frac{d\xi}{\gamma(\xi)^\frac{1}{c}}
\oint_{ {\tilde{\gamma}_{2}}\times {\tilde{\gamma}}_{3}\times\gamma_{4}\times\gamma_{5}}
\frac{\partial_{\xi}J_{1}}{J_{1}}\frac{ {x}_{1}\,d {x}_{2}d {x}_{3}
d {x}_{4}d {x}_{5}}{P}\\
=&\int_{\xi_{0}}^{\xi}\frac{d\xi}{\gamma(\xi)^\frac{1}{c}}
\oint_{\gamma_{1}\times {\tilde{\gamma}}_{2}\times {\tilde{\gamma}}_{3}\times\gamma_{4}
\times\gamma_{5}}\frac{\partial_{\xi}J_{1}}{J_{1}}\frac{d {x}_{1}d {x}_{2}d
 {x}_{3}d {x}_{4}d {x}_{5}}{P}\nonumber\ .
\end{align}
In the first line of the above expression, we have specified the two-cycle $C$ by considering two small circles $ {\tilde{\gamma}}_{2}$ and $ {\tilde{\gamma}}_{3}$ which encircle $J_{0}=0$ and $J_{2}=0$ respectively. We notice that $ {x}_{4}$ is given by the solution of $J_{1}=0$, and for sufficiently small $\xi$ in the vicinity of the LG-point, $ {x}_{4}$ tends to zero. Therefore, $\gamma_{4}$ encircles $ {x}_{4}$ at origin, and the value of the  $d {x}_{4}$ integral is given by taking the residue of the integrand around $ {x}_{4}=0$. In the second line of (\ref{sup4}), $\gamma_{5}$ encircles $ {x}_{5}$ at origin and we have used the fact that $\frac{1}{2\pi i}\oint_{\gamma_{5}} \frac{d {x}_{5}}{P}=\Big(\frac{\partial P}{\partial  {x}_{5}}\Big)^{-1}$. It is important to note that as $J_{0}\rightarrow0$ and $J_{2}\rightarrow0$, the coordinate $ {x}_{5}$ which is given by the solution of $P=0$ tends to zero, for sufficiently small values of $\psi$. Therefore, the value of $d {x}_{5}$ integral is given by the residue of the integrand evaluated around $ {x}_{5}=0$. In the last line of (\ref{sup4}), we used the fact that $ {x}_{1}$ is constant and we have inserted the identity $\frac{1}{2\pi i}\oint_{\gamma_{1}}\frac{d {x}_{1}}{ {x}_{1}}=1$ in which $\gamma_{1}$ encircles $ {x}_{1}$ at the origin. Of course, we have dropped all overall constant factors at the end.
\par
In the next step, in order to get closer to the simplified Kapustin-Li formula, we perform the following change of variables
\begin{eqnarray}\label{chvar2}
 {x}_{1}\mapsto  {x}_{1}\quad,\quad  {x}_{2}\mapsto v=J_{0}( {x})\quad,\quad  {x}_{3}\mapsto w=J_{2}( {x})\quad,\quad  {x}_{4}\mapsto  {x}_{4}\quad,\quad  {x}_{5}\mapsto  {x}_{5}\ .
\end{eqnarray}
We notice that to simplify the Kapustin-Li formula, we made the same change of variable (\ref{chvar}). Under (\ref{chvar2}), we find for the superpotential (\ref{sup4})
\begin{eqnarray}\label{sup5}
{\cal{W}}(\psi,\xi)=\int_{\xi_{0}}^{\xi}\frac{d\xi}{\gamma(\xi)^\frac{1}{c}}\oint_{\gamma_{1}
\times\gamma_{2}\times\gamma_{3}\times\gamma_{4}\times\gamma_{5}} \frac{\partial_{\xi}J_{1}}{J_{1}}
\frac{dv dw\,d {x}_{1}d {x}_{4}d {x}_{5}}{\partial_{ {x}_{2}}v\,
\partial_{ {x}_{3}}w\,P},
\end{eqnarray}
where $\gamma_{2}$ and $\gamma_{3}$ encircle $v$ and $w$ coordinates at their origins respectively. In other words, after imposing (\ref{chvar2}), to evaluate the $dv$ and $dw$ integrals in (\ref{sup5}), one needs to compute the residue of the integrand at $v=0$ and $w=0$. Now, let us split the defining equation of the Calabi-Yau hypersurface $Y$ into the Fermat polynomial and the deformation terms, i.e. $P(x;\psi)=W(x)-\psi\,G(x)$ ($W(x)$ is the Fermat polynomial and $G(x)$ is a deformation added around the Fermat point). Since we are calculating the D2-brane superpotential in the vicinity of the LG-point in moduli space, we can expand (\ref{sup5}) in the region of small $\psi$ in the following way
\begin{eqnarray}\label{supexp}
{\cal{W}}(\psi,\xi)=\sum_{n=0}^{\infty}\psi^{n}\int_{\xi_{0}}^{\xi}\frac{d\xi}
{\gamma(\xi)^\frac{1}{c}}\oint_{\times_{i=1}^{5}\gamma_{i}} \frac{\partial_{\xi}J_{1}}{J_{1}}\Big(\frac{G}{W}\Big)^{n}\,\,
\frac{dv dw\,d {x}_{1}d {x}_{4}d {x}_{5}}{\partial_{ {x}_{2}}v\,
\partial_{ {x}_{3}}w\,W}\ .
\end{eqnarray}
We immediately notice that in (\ref{supexp}), we do not gain any contribution from $n=0$. The only way that one can find a non-vanishing contribution from $n=0$ is if the Fermat polynomial contains the monomial $v\,w\, {x}_{1} {x}_{4} {x}_{5}$. However, since $J_{0}$ only involves $x_{1}$ and $x_{2}$, and $J_{2}$ only contains $x_{3}$, $x_{4}$ and $x_{5}$, it is impossible for $v$ and $w$ to get mixed in the Fermat polynomial $W$ under the change of coordinates (\ref{chvar2}). Therefore, we arrive at
\begin{eqnarray}\label{sup6}
{\cal{W}}(\psi,\xi)=\sum_{n=1}^{\infty}\psi^{n}\int_{\xi_{0}}^{\xi}\frac{d\xi}
{\gamma(\xi)^\frac{1}{c}}\oint_{\times_{i=1}^{5}\gamma_{i}} \frac{\partial_{\xi}J_{1}}{J_{1}}\Big(\frac{G}{W}\Big)^{n}\,\,
\frac{dv dw\,d {x}_{1}d {x}_{4}d {x}_{5}}{\partial_{ {x}_{2}}v\,
\partial_{ {x}_{3}}w\,W}\ .
\end{eqnarray}
As is clear, the above superpotential (\ref{sup6}) is exact in both closed- and open-string moduli. However, the D-brane superpotential we found in the previous section from the Kapustin-Li formula is exact in the open-string modulus, but first order in the closed-string modulus. To make a comparison between the two formulas, we first have to pick the first closed-string order result of (\ref{sup6})
\begin{eqnarray}\label{suplin}
{\cal{W}}^{(n=1)}(\psi,\xi)=\psi\int_{\xi_{0}}^{\xi}\frac{d\xi}{\gamma(\xi)^\frac{1}{c}}
\oint_{\times_{i=1}^{5}\gamma_{i}} G\,\frac{\partial_{\xi}J_{1}}{J_{1}}\,
\frac{dv dw\,d {x}_{1}d {x}_{4}d {x}_{5}}{\partial_{ {x}_{2}}v\,
\partial_{ {x}_{3}}w\,W^{2}}\ .
\end{eqnarray}
We again notice that the only way to obtain a non-vanishing contribution for the above superpotential is to have monomials in the denominator of the integrand of (\ref{suplin}), which necessarily mix $v$ and $w$ coordinates with certain powers (the right powers are determined by the monomials which appear in the deformation term $G$). However, as we discussed earlier, $v$ and $w$ do not get mixed in the Fermat polynomial. Therefore, the only way to find monomials which include both $v$ and $w$ from $W^{2}$ is to have the $v$ dependence from one of the $W$ factors and the $w$ dependence from the other $W$ factor. It is clear that the whole $v$ dependence of $W$ comes from $x_{2}^{A_{2}}$ through the change of variables (\ref{chvar2}), and similarly, the $w$ dependence of $W$ is encoded in $x_{3}^{A_{3}}$ through (\ref{chvar2}). Hence, we can substitute $x_{2}^{A_{2}}$ by $\frac{1}{A_{2}}v\,\partial_{v}W$ and $x_{3}^{A_{3}}$ by $\frac{1}{A_{3}}w\,\partial_{w}W$ after imposing (\ref{chvar2}). Dropping the overall numerical constant, we obtain for the first closed-string order D2-brane superpotential \begin{eqnarray}\label{sup7}
{\cal{W}}^{(n=1)}(\psi,\xi)=\psi\int_{\xi_{0}}^{\xi}\frac{d\xi}{\gamma(\xi)^\frac{1}{c}}
\oint_{\times_{i=1}^{5}\gamma_{i}} G\,\frac{\partial_{\xi}J_{1}}{J_{1}}\,
\frac{dv dw\,d {x}_{1}d {x}_{4}d {x}_{5}}{\partial_{ {x}_{2}}v\,
\partial_{ {x}_{3}}w\,v\,\partial_{v}W\,w\,\partial_{w}W}\ ,
\end{eqnarray}
which agrees precisely  with the simplified Kapustin-Li formula (\ref{com}). As expected, both geometric and worldsheet approaches give rise to the same D2-brane superpotential for the 4-dimensional ${\cal{N}}=1$ physics.

Before closing this section, let us point out an interesting observation. By constructing the three-chain $\Gamma_{0}$ whose base two-cycle is defined via (\ref{Ccyc}), we conjecture that (\ref{sup5}) captures the full dependence of the effective superpotential on both open- and closed-sting moduli. Having the exact result, it is clear that (\ref{sup6}) represents a perturbative expansion of the effective superpotential in terms of the closed-string modulus. Consequently, one can, in principle, compute the superpotential at an arbitrary order $n$ by inserting the factor $\big(\frac{G}{W}\big)^{n}$ in the integrand of the integral and evaluating the residues. It would be very interesting to derive this result from the first principles of the worldsheet obstruction and matrix factorization techniques for this class of models.

\section{An Explicit Example}
\setcounter{equation}{0}
As an explicit example, in this section we choose a particular compact Calabi-Yau hypersurface equipped with a D2-brane open sector. We then proceed and compute the superpotential associated with the deformations of this D-brane configuration, both from the worldsheet and geometric point of view. As we observed in the general proof presented in the last section, from the worldsheet point of view, we can capture the first closed-string order superpotential, whereas the geometric chain integral gives us the exact result. Let us first start by introducing the B-model bulk geometry. The bulk Calabi-Yau geometry is chosen to be a quasi-homogeneous degree eight polynomial in the  weighted projective space ${\mathbb{P}}^{4}_{(1,1,2,2,2)}$. The defining equation of this Calabi-Yau, $Y={\mathbb{P}}^{4}_{(1,1,2,2,2)}[8]$, at the Fermat point is given by
\begin{eqnarray}\label{ferW}
W=x_{1}^{8}+x_{2}^{8}+x_{3}^{4}+x_{4}^{4}+x_{5}^{4}\ ,
\end{eqnarray}
which develops an $A_{1}$ singularity along the loci $x_{1}=x_{2}=0$. This Calabi-Yau space has the Hodge numbers $h^{1,1}(Y)=2$ and $h^{2,1}(Y)=86$. Therefore, in the most general case, one is allowed to deform $W$ in 86 different ways by adding appropriate monomials to it. However, only a subset of these perturbations does not excite the unobstructed fermions. In the sequel, after introducing the geometry of the open sector, we will restrict ourselves to the afore slices of the whole complex structure moduli space of the above Calabi-Yau $Y$, and we will then compute the D-brane superpotential induced by these deformations in a chosen branch of the open moduli space.

The moduli space of D2-branes in this Calabi-Yau consists of several branches which may intersect in finitely many points. Any such family of branes is described by embedding a ${\mathbb{P}^{1}}$ into the ambient projective space, ${\mathbb{P}}^{1}\hookrightarrow{\mathbb{P}^{4}}:\,\,(u:v)\mapsto (x_1: x_2: x_3: x_4: x_5)$. For Fermat type Calabi-Yaus such mappings fall into distinct classes, where each class is determined by the choice of a pair of coordinates.
For any pair $(x_i, x_j)$ we can make an ansatz for its parametrization by setting $x_i=u^m$ and $x_j=\chi u^n$, where $n$ and $m$ are suitable exponents which are generally different from one in weighted projective spaces. The fact that the embedded ${\mathbb{P}^{1}}$ is part of the Calabi-Yau determines $\chi$, as well as another constraint which must be fulfilled by the remaining coordinates.

The number of different branches of the moduli space of the embedding ${\mathbb{P}^{1}}$ is determined by the possible permutation symmetries of the coordinates of the ambient Calabi-Yau space. In the most general case, there are 10 branches for a Fermat three-fold \cite{Baumgartl:2008qp}.  In the example (\ref{ferW}), which we study here, we can identify three different types of branches, specified by choosing the coordinates of the embedded ${\mathbb{P}^{1}}$ as $(x_1, x_2)$, $(x_1, x_3)$ and $(x_3, x_4)$. In the following we will analyze each of these branches separately.

\subsection{Branch I of the Open Moduli Space}
For the first branch of the open moduli space, we embed a family of ${\mathbb{P}^{1}}$'s into the ambient Calabi-Yau by choosing the pair $(x_1, x_2)$. Since these two coordinates have the same weight, they satisfy a linear relation
\eqn{\label{embtwo}
	(x_1:x_2:x_3:x_4:x_5) &= (u: \eta u: \alpha v: \beta v: \gamma v) \ .
}
The remaining three coordinates have weight 2, and therefore, the moduli space of this family of holomorphic curves is given as a Riemann surface described as a hypersurface in ${\mathbb P}^2$
\eqn{\label{msp1}
	\alpha^4+\beta^4+\gamma^4 = 0 \qquad,\qquad \eta^8=-1\ .
}
This moduli space (${\mathbb{P}^{2}}[4]$) parameterizes a family of complex lines whose associated matrix factorization is constructed in below. To calculate the superpotential on this branch, we consider bulk deformation operators which excite the single exactly marginal fermion. In this way, the correlation functions have an interpretation as holomorphic differentials on the moduli space. The Riemann surface (\ref{msp1}) has genus 3. Therefore, we expect to find three corresponding deformations which generate the effective superpotential on this branch. In fact, these deformations only take care of $x_{3}$, $x_{4}$, and $x_{5}$ fields which appear in the deformation operators. In addition, we must consider the multiplicity of the $x_{1}$ and $x_{2}$ fields that can appear in the deformation terms. The fact that the total charge of the deformation operator is two fixes this multiplicity factor to 7. Therefore, we find the total number of 21 deformation operators. Finally, in order to find the effective superpotential, we need to integrate these correlators with respect to the open modulus.

The lines (\ref{embtwo}) have an associated family of matrix factorizations. One explicit representation is given by the following choice of polynomials:
\eqn{
	J_0 &=  x_2-\eta x_1 \\
	E_0 &=  x_2^7+\eta x_1 x_2^6+\eta^2 x_1^2 x_2^5+\eta^3 x_1^3 x_2^4+\eta^4 x_1^4 x_2^3+\eta^5 x_1^5 x_2^2+\eta^6 x_1^6 x_2+\eta^7 x_1^7 \\
	J_1 &= \alpha x_4-\beta x_3 \\
	E_1 &= \frac {x_4^3}{\alpha}+\frac{\beta x_3 x_4^2}{\alpha^2}+\frac{\beta^2 x_3^2 x_4}{\alpha^3}+\frac{\beta^3 x_3^3}{\alpha^4} \\
	J_2 &= \alpha x_5-\gamma x_3 \\
	E_2 &= \frac {x_5^3}{\alpha}+\frac{\gamma x_3 x_5^2}{\alpha^2}+\frac{\gamma^2 x_3^2 x_5}{\alpha^3}+\frac{\gamma^3 x_3^3}{\alpha^4} \ .
}
The complex structure deformations which generate an effective superpotential on this branch without excitation of obstructed fermions in the marginal cohomology have the following form
\eqn{
	G &= r^{(6)}(x_1, x_2) s^{(2)} (x_3, x_4, x_5)
		= r^{(6)}(x_1, x_2) ( s^{(2)}_{100} x_3 + s^{(2)}_{010} x_4 + s^{(2)}_{001} x_5)\ .
}
Here $s^{(n)}$ is a homogenous polynomial of degree $n$. It is straightforward to compute the correlation function between the bulk deformation operator and the exactly marginal open string fermion:
\eqn{
	\langle G \partial_\beta Q \rangle
	=-\frac{\eta \alpha r^{(6)}(1, \eta)}{32}\frac{s^{(2)}(\alpha,\beta, \gamma)}{\gamma^3} \ .
}
Integrating the above F-term gives rise to the first order contribution to the effective superpotential on this branch
\eqn{
\label{branch1res}
	{\cal W} &= -\frac{\eta r^{(6)}(1, \eta)}{32}
		\sum_{abc}s^{(2)}_{abc} \frac{\eta^{10c+2}\beta^{1+b}}{b+1} {}_2F_1(\tfrac{b+1}{4}, \tfrac{3-c}{4}; \tfrac{5+b}{4}; -\beta^4)\ .
}

Now, we would like to compute the D2-brane superpotential on this branch of the open moduli space from the geometric perspective. One way of obtaining the superpotential is to directly compute the chain integral (\ref{sup3}) with the two-cycle $C$ defined in (\ref{Ccyc}). Although it is possible to perform this direct computation, we find it easier to first compute one of the relative periods of the system in a mixed phase of the moduli space via the technique of direct integration \cite{Jockers:2009mn}. Then, we analytically continue this solution to the vicinity of the LG-point in open/closed moduli space and, as a result, we generate a set of solutions to the open/closed Picard-Fuchs equations in the vicinity of the LG-point. A priori, the D2-brane superpotential is a linear combination of these solutions. Since we have already computed the first closed-string order superpotential from the worldsheet perspective, in order to identify the exact superpotential, we use the first order result as a guide to choose the right linear combination among the solutions.
\par
Let us point out that another way of finding the relative periods of the open/closed geometry in the vicinity of the LG-point, as is shown in \cite{Aganagic:2009jq}, is to establish a system of linear differential equations whose solutions are relative periods. However, since for this class of examples, one does not necessarily consider the whole set of possible bulk deformations, one finds differential operators whose ranks are greater than four \cite{Aganagic:2009jq}. As a consequence, this system of differential equations contains many unphysical solutions that should be disregarded. It is also possible to generate this system of differential equations by the techniques of \cite{Alim:2009rf} and we hope to come back to this problem and other related issues in a future work \cite{AHJMMS}.
\par
To compute the D2-brane effective superpotential on this branch of the moduli space, we start with the defining equation of the deformed Calabi-Yau $Y$
\begin{eqnarray}\label{bulkg}
P=W-r^{(6)}(x_{1},x_{2})\,x_{3}\ ,
\end{eqnarray}
in which we have chosen $s^{(2)}(x_{3},x_{4},x_{5})=x_{3}$, and we have kept $r^{(6)}(x_{1},x_{2})$ arbitrary at the moment. According to the procedure of $N=1$ special geometry, in the next step we need to introduce a family of divisors which embeds the family of holomorphic curves (\ref{embtwo}). To do this, we simply pick
\begin{eqnarray}\label{divtwo}
Q(\xi)=x_{5}-\xi\,x_{4}\ ,
\end{eqnarray}
where $\xi$ is the modulus of the family of divisors. We notice that for any given member of the family of holomorphic two-cycles (\ref{embtwo}), there is a corresponding divisor in the family (\ref{divtwo}) with the modulus $\xi=\frac{\gamma}{\beta}$, which encompasses the given holomorphic two-cycle. Once we have introduced the geometry of the open sector, we can construct the holomorphic two-form of the subsystem
\begin{eqnarray}\label{holtwotwo}
\partial_{\xi}\underline{\Omega}\simeq-\int\frac{x_{4}}{P\, Q(\xi)}\Delta\ ,
\end{eqnarray}
where $\Delta=\sum_{i}a_{i}\,x_{i}dx_{1}\wedge\cdots\widehat{dx_{i}}\cdots\wedge dx_{5}$. Now, if we impose the following change of variables in the integrand
\begin{eqnarray}\label{chvartwo}
y_{1}=x_{1}\quad,\quad y_{2}=x_{2}\quad,\quad y_{3}=x_{3}\quad,\quad y_{4}=(1+\xi^{4})^{1/2}\,x_{4}^{2}\quad,\quad y_{5}=x_{5}-\xi\,x_{4}\ ,
\end{eqnarray}
on (\ref{holtwotwo}), and perform the residue integral along the contour which encircles $y_{5}$ coordinate at its origin, we find the transformed holomorphic two-form of the subsystem to be
\begin{eqnarray}\label{tranholtwo}
\partial_{\xi}\underline{\Omega}\simeq\frac{1}{(1+\xi^{4})^{1/2}}\int\frac{1}{\tilde{P}}
\Delta_{y}\ ,
\end{eqnarray}
where $\tilde{P}$ and $\Delta_{y}$ are given by
\begin{eqnarray}\label{K3two}
\tilde{P}=y_{1}^{8}+y_{2}^{8}+y_{3}^{4}+y_{4}^{2}-s^{(6)}(y_{1},y_{2})y_{3}\quad,\quad
\Delta_{y}=\sum_{i=1}^{4}\tilde{a}_{i}\,y_{i}dy_{1}\wedge\cdots\widehat{dy_{i}}\cdots\wedge dy_{4}\ .
\end{eqnarray}
We easily recognize that, as far as the periods are concerned, the subsystem periods of the above open/closed geometry coincide with the periods of the K3 surface ${\mathbb{P}}^{3}_{(1,1,2,4)}[8]$ which is defined by the equation $\tilde{P}=0$ in the weighted projective space ${\mathbb{P}}^{3}_{(1,1,2,4)}$.
\par
Moreover, we realize that the whole dependence on $\xi$ in (\ref{tranholtwo}) is now encoded in the prefactor of the integral resulting from the Jacobian of the coordinate transformation. Therefore, integration over a two-cycle of the subsystem only contributes to the closed-string dependence of the subsystem periods, and the open-string dependence of the periods is entirely captured by the prefactor. Hence, after integrating over a two-cycle of the subsystem followed by an integration over the open modulus $\xi$, the periods of the open/closed system are found to have the following structure
\begin{eqnarray}\label{depen}
\Pi_{a}(\xi,\psi_{i})=f(\xi)\,g_{a}(\psi_{i})\quad,\quad\mbox{where \ \ }f(\xi)=\xi\,
{ }_{2}F_{1}(\tfrac{1}{4},\tfrac{1}{2};\tfrac{5}{4};-\xi^{4})\ ,
\end{eqnarray}
and $\psi_{i}$ is the closed-string modulus associated with a closed-string deformation around the Fermat point. To be concrete, let us now make a particular choice for $r^{(6)}(x_{1},x_{2})$ and fully compute the periods. In principle, $r^{(6)}$ can be any homogeneous polynomial of degree 6 in terms of $x_{1}$ and $x_{2}$. But for simplicity, let us pick the monomial $r^{(6)}(x_{1},x_{2})=\psi\,x_{1}^{4}x_{2}^{2}$ among many other allowed ones. Therefore, the defining equation of the subsystem, $\tilde{P}$, reads
\begin{eqnarray}\label{Ppsi}
\tilde{P}(\psi)=y_{1}^{8}+y_{2}^{8}+y_{3}^{4}+y_{4}^{2}+\psi\,y_{1}^{4}
y_{2}^{2}y_{3}\ .
\end{eqnarray}
In order to take the advantage of the direct integration method {\textit{at a large complex structure limit}} \cite{Jockers:2009mn}, we add an auxiliary perturbation $\lambda\,y_{1}y_{2}y_{3}y_{4}$ to $\tilde{P}$ and compute the corresponding periods in the limit of large $\lambda$. Since we find the result as an analytic function in terms of moduli $\psi$ and $\lambda$, we can then analytically continue the result to the region of small $\lambda$, and finally set $\lambda=0$. Therefore, as we just described, first we would like to compute
\begin{eqnarray}\label{perG}
G(\psi,\lambda)=\int_{\times_{i=1}^{4}\sigma_{i}}\frac{dy_{1}\,dy_{2}\,dy_{3}\,dy_{4}}
{\tilde{P}(\psi)-\lambda\,y_{1}y_{2}y_{3}y_{4}}\ ,
\end{eqnarray}
in the limit of large $\lambda$. In (\ref{perG}), $\sigma_{i}$ is the contour which encircles $y_{i}$ coordinate at its origin. We should notice that the two-cycle (specified by $\sigma_{i}$'s) in the subsystem we are integrating over in this case is clearly a different cycle from (\ref{Ccyc}). However, at the end, we will be able to choose the period which corresponds to $C$. Following the method of \cite{Jockers:2009mn}, we arrive at
\begin{eqnarray}\label{perG2}
G(\psi,\lambda)=-\frac{1}{\lambda}\sum_{n=0}^{\infty}\sum_{k=0}^{[\frac{n}{2}]}
\frac{(8n)!}{(4k)!\,(4n)!\,(n-2k)!\,(n-k)!\,(2n-k)!}\frac{\psi^{4k}}{\lambda^{8n}}\ .
\end{eqnarray}
The above expression is convergent in the vicinity of small $\psi$ and large $\lambda$. In order to evaluate (\ref{perG2}) at $\lambda=0$, we first need to analytically continue this expression to the region of small $\lambda$. To achieve this, we recognize that (\ref{perG2}) has the following Barnes integral representation
\begin{eqnarray}\label{Barnes1}
G(\psi,\lambda)=-\frac{1}{2\pi i}\sum_{k=0}^{\infty}\oint_{B}d\nu\,
\frac{(-1)^{\nu}\,\Gamma(-\nu)\Gamma(\nu+1)\Gamma(8\nu+1)}{(4k)!\,\Gamma(4\nu+1)
\Gamma(\nu-2k+1)\Gamma(\nu-k+1)\Gamma(2\nu-k+1)}\frac{\psi^{4k}}{\lambda^{8\nu+1}}\ .
\end{eqnarray}
If one closes the contour $B$ from the right, one restores (\ref{perG2}) from the   contributions of the simple poles on the positive side of the real axis. However, if one closes the contour $B$ from the left, one obtains the analytic expression for $G(\psi,\lambda)$ in the vicinity of small $\lambda$ and $\psi$. We recognize that upon the closure of the contour from the left, $B$ encompasses four sets of simple poles located at $\nu=-m-a/8$ where $m=0,1,2,\cdots$ and $a\in\{1,3,5,7\}$. Since we eventually want to set $\lambda=0$, the whole contribution to (\ref{Barnes1}) at $\lambda=0$ comes from the simple pole $\nu=-1/8$. Evaluating the residue of (\ref{Barnes1}) at $\nu=-1/8$ and setting $\lambda=0$, we find a hypergeometric function in terms of $\psi$. Once we have a hypergeometric function as the solution to the differential equations of the open/closed system, we can easily construct the other solutions of the system. For the case at hand, we find the following four solutions
\begin{align}
g_{1}(\psi)&={ }_{3}F_{2}\Big(\frac{1}{16},\frac{9}{16},\frac{1}{8};\frac{2}{4},\frac{3}{4};
\frac{\psi^{4}}{2^{6}}\Big)\ ,\label{gone}\\
g_{2}(\psi)&=\psi\,{ }_{3}F_{2}\Big(\frac{5}{16},\frac{13}{16},\frac{3}{8};\frac{3}{4},\frac{5}{4};
\frac{\psi^{4}}{2^{6}}\Big)\ ,\label{gtwo}\\
g_{3}(\psi)&=\psi^{2}\,{ }_{3}F_{2}\Big(\frac{9}{16},\frac{17}{16},\frac{5}{8};\frac{5}{4},\frac{6}{4};
\frac{\psi^{4}}{2^{6}}\Big)\ ,\label{gthr}\\
g_{4}(\psi)&=\psi^{3}\,{ }_{4}F_{3}\Big(\frac{13}{16},\frac{21}{16},\frac{7}{8},1;\frac{5}{4},\frac{6}{4}
,\frac{7}{4};\frac{\psi^{4}}{2^{6}}\Big)\ .\label{gthr}
\end{align}
It is clear from above solutions that only (\ref{gtwo}) starts with a linear piece in terms of the closed-string modulus $\psi$. Therefore, due to (\ref{depen}), the superpotential on this branch of the moduli space reads
\begin{eqnarray}\label{supbr1}
{\cal W}(\psi,\xi)=\psi\,\xi\,{ }_{2}F_{1}\Big(\frac{1}{4},\frac{1}{2};\frac{5}{4};-\xi^{4}\Big)\,{ }_{3}F_{2}\Big(\frac{5}{16},\frac{13}{16},\frac{3}{8};\frac{3}{4},\frac{5}{4};
\frac{\psi^{4}}{2^{6}}\Big)\ .
\end{eqnarray}
Although (\ref{supbr1}) is a purely classical superpotential of a D2-brane configuration in the B-model, it contains highly nontrivial information about the non-perturbative superpotential of the corresponding Lagrangian brane in the mirror A-model. To extract the A-model superpotential, one first needs to identify the correct open mirror map. Having the right mirror map, one can then construct the A-model superpotential by applying the open mirror map to the classical B-model superpotential expressed in {\textit{flat coordinates}}. It has been shown in \cite{Alim:2009bx} that the flatness of the Gauss-Manin connection defined on the open/closed moduli space determines the correct open- and closed-string mirror maps. However, in order to apply this criterion to specify the distinguished open mirror map, one first needs to have a well-defined notion of Gauss-Manin connection defined on the open/closed moduli space. For the class of models under consideration in this paper, this is a nontrivial task to do. We postpone a detailed study of the Gauss-Manin system and further connections to the A-model superpotential for this class of models for a future work \cite{AHJMMS}.


Now, let us compare the results of the two approaches. To to the comparison, we first notice that the chosen perturbation (\ref{bulkg}) corresponds to the contribution of $s^{(2)}_{100}$ in (\ref{branch1res}). Second, as mentioned earlier, the modulus associated with the family of divisors, $\xi$, is related to the modulus of holomorphic curves (\ref{embtwo}) by $\xi=\frac{\gamma}{\beta}$. Using (\ref{msp1}), it is easily seen that $\beta$ and $\xi$ are related via
\eqn{
	-\beta^4 &= \frac{1}{1+\xi^4}\ ,
}
in a patch where $\alpha=1$. In order to compare the two hypergeometric functions appearing in (\ref{branch1res}) and (\ref{supbr1}), it is necessary to perform suitable  transformations on the hypergeometric functions. We impose two successive transformations on the argument of the hypergeometric function in (\ref{supbr1}). We perform an analytic continuation first from $z\to\frac{1}{z}$ and then followed by   $z\to\frac{z}{z-1}$. Under these transformations, we have the following identities for  a hypergeometric function ${ }_{2}F_{1}(a,b;c;z)$ when $a-b$ and $c-a-b$ are not integers \cite{Abramowitz:1965aa}
\eqn{\label{hypid}
	{}_2F_1(a,b;c;z) &=\frac{ \Gamma(c)\Gamma(b-a) }{\Gamma(b)\Gamma(c-a)} (-z)^{-a} \;{}_2F_1\Bigl(a, 1-c+a; 1-b+a; \frac{1}{z}\Bigr)
		+ (a \leftrightarrow b) \ ,\\
{}_2F_1(a,b;c;z) &=(1-z)^{-a}{ }_{2}F_{1}\Bigl(a,c-b;c;\frac{z}{z-1}\Bigr)=(1-z)^{-b}{ }_{2}F_{1}\Bigl(b,c-a;c;\frac{z}{z-1}\Bigr)\ .
}

These two transformations bring the argument of the hypergeometric function (\ref{supbr1}) in the correct form. Thus, we have
\eqn{
\label{B1MFres}
	\xi\; {}_2F_1(\tfrac{1}{4},\tfrac{1}{2};\tfrac{5}{4};-\xi^4)
	&= \frac{\Gamma(\frac{5}{4})\Gamma(\frac{1}{4})}{\Gamma(\frac{1}{2})\Gamma(\frac{3}{4})}
		-
\frac{1}{\xi}\; {}_2F_1(\tfrac{1}{2},\tfrac{1}{4};\tfrac{5}{4};-\xi^{-4})
	= \text{const.} - \eta^2 \beta
			\; {}_2F_1(\tfrac{1}{4},\tfrac{3}{4};\tfrac{5}{4};-\beta^4) \ .
}

%
%
%
%
Comparison of (\ref{B1MFres}) and (\ref{supbr1}) with the correct integration domains  taken into account shows that the contributions to the effective superpotentials computed in two different ways are identical
\eqn{
	\xi\; {}_2F_1(\tfrac{1}{4},\tfrac{1}{2};\tfrac{5}{4};-\xi^4) \Bigr|_{\xi_0}^\xi
	\propto \beta
			\; {}_2F_1(\tfrac{1}{4},\tfrac{3}{4};\tfrac{5}{4};-\beta^4) \Bigr|_{\beta_0}^\beta\ .
}
The relative overall constant can be reabsorbed in the normalization of the holomorphic three-form of the Calabi-Yau.

\subsection{Branch II of the Open Moduli Space}
The second branch of the open moduli space is parameterized by picking the pair $(x_1, x_4)$ as the coordinates of the embedded ${\mathbb{P}^{1}}$. Thus, we have
\eqn{\label{famhol}
	(x_1:x_2:x_3:x_4:x_5) &= (u: \alpha v:  \beta v^2: \eta^2 u^2: \gamma v^2) \ .
}
The moduli space associated with the above family of holomorphic curves is a Riemann surface realized as a hypersurface in ${\mathbb P}^{2}_{(122)}$. The defining hypersurface equation of ${\mathbb P}^{2}_{(122)}[8]$ is given by
\begin{eqnarray}\label{mosp2}
\alpha^{8}+\beta^{4}+\gamma^{4}=0\qquad,\qquad \eta^{8}=-1\ .
\end{eqnarray}
This Riemann surface has genus 3. One conveniently finds a matrix factorization associated with the family (\ref{famhol}). The factorization is explicitly given by
\eqn{
	J_0 &= x_4-\eta^2 x_1^2 \\
	E_0 &= x_4^3+\eta^2 x_4^2 x_1^2+\eta^4 x_1^4 x_4+\eta^6 x_1^6 \\
	J_1 &= \alpha^2 x_3-\beta x_2^2 \\
	E_1 &= \frac{x_3^3}{\alpha^2}+\frac{\beta x_2^2 x_3^2}{\alpha^4}+\frac{\beta^2 x_2^4 x_3}{\alpha^6}+\frac{\beta^3 x_2^6}{\alpha^8} \\
	J_2 &= \alpha^2 x_5-\gamma x_2^2 \\
	E_2 &= \frac{x_5^3}{\alpha^2}+\frac{\gamma x_2^2 x_5^2}{\alpha^4}+\frac{\gamma^2 x_2^4 x_5}{\alpha^6}+\frac{\gamma^3 x_2^6}{\alpha^8} \ .
}
We now deform the bulk geometry by considering operators which have a correlation function with exactly one marginal fermion. There are 9 such deformations and they are  given by
\eqn{\label{deforG}
	G &= r^{(5)}(x_1, x_4) s^{(3)}(x_2, x_3, x_5)\ .
}
The above deformations lift the flat modulus $\beta$ and generate an effective superpotential. The bulk-boundary correlation function is easily computed as
\eqn{
	\langle G \partial_\beta Q \rangle
	= -\frac{1}{16} \alpha^2\eta^2 r^{(5)}(1,\eta^2)\frac{s^{(3)}(\alpha, \beta, \gamma)}{\gamma^3}\ ,
}
which results in the superpotential
\eqn{
\label{MFsup2}
	{\cal W} &= -\frac{1}{16}\eta^2 r^{(5)}(1,\eta^2)
		\sum_{abc}s^{(3)}_{abc} \frac{\eta^{10c+2}\beta^{1+b}}{b+1} {}_2F_1(\tfrac{b+1}{4}, \tfrac{3-c}{4}; \tfrac{5+b}{4}; -\beta^4) \ .
}

\par
As in branch I, we would like to compute the effective D2-brane superpotential on this branch of the moduli space from the geometric perspective as well. According to (\ref{deforG}), the defining equation of the Calabi-Yau is given by
\begin{eqnarray}\label{PCY}
P=W+r^{(5)}(x_{1},x_{4})
s^{(3)}(x_{2},x_{3},x_{5})\ .
\end{eqnarray}
It is clear that the most general form for $s^{(3)}$ is $s^{(3)}(x_{2},x_{3},x_{5})=\psi\,x_{2}x_{3}+\phi\,x_{2}x_{5}$, where $\psi$ and $\phi$ are two closed-string moduli. For $r^{(5)}$, there are three possible monomials $\{x_{1}^{5},x_{1}^{3}x_{4},x_{1}x_{4}^{2}\}$ that one can consider, and for simplicity, we choose only one of them $r^{(5)}(x_{1},x_{4})=x_{1}^{3}x_{4}$.
\par
To define the open sector of the geometry, we need to introduce a family of divisors which encompasses the given family of holomorphic curves (\ref{famhol}). In this case, we choose this family to be
\begin{eqnarray}\label{div2}
Q(\xi)=x_{5}-\xi\,x_{3}\ ,
\end{eqnarray}
where $\xi$ is the open-string modulus. Given a member of the family of holomorphic curves in (\ref{famhol}), the family of divisors encompasses the given curve with the value $\xi=\frac{\gamma}{\beta}$ for the open modulus. Performing the following change of coordinates
\begin{eqnarray}\label{chcoor}
y_{1}=x_{1}\quad,\quad y_{2}=x_{2}\quad,\quad y_{3}=(1+\xi^{4})^{1/4}x_{3}\quad,\quad y_{4}=x_{4}\quad,\quad y_{5}=x_{5}-\xi\,x_{3}\ ,
\end{eqnarray}
we easily recognize that the holomorphic two-form of the subsystem of the open/closed geometry transforms as
\begin{eqnarray}\label{twoform}
\partial_{\xi}\underline{\Omega}\simeq-\int\frac{x_{3}}{P(\psi,\phi)\,Q(\xi)}\,\Delta
\simeq\frac{1}{(1+\xi^{4})^{1/2}}\int\frac{y_{3}}{\tilde{P}(\rho)}\,\Delta_{y}\ .
\end{eqnarray}
In the last expression in (\ref{twoform}), we have performed the residue integral on the $y_{5}$ coordinate. Moreover, $\tilde{P}(\alpha)$ and $\alpha$ in (\ref{twoform}) are defined as
\begin{eqnarray}\label{trantwofor}
\tilde{P}(\rho)=y_{1}^{8}+y_{2}^{8}+y_{3}^{4}+y_{4}^{4}
+\rho\,y_{1}^{3}y_{2}y_{3}y_{4}\qquad,\qquad \rho=\frac{\psi+\phi\,\xi}{(1+\xi^{4})^{1/4}}\ .
\end{eqnarray}
From the definition of $\tilde{P}$, we realize that the periods of the subsystem are identically the periods of the complex surface ${\mathbb{P}}^{3}_{(1,1,2,2)}[8]$ which is not a K3 surface. This can be also seen from the holomorphic two-form of the subsystem, (\ref{twoform}), by realizing the fact that because of the presence of the $y_{3}$ factor in the numerator, this form is not a nowhere vanishing holomorphic two-form, which is required for a K3 surface. Nonetheless, we show that we can still apply the technique of direct integration to compute the periods of the holomorphic two-form of the subsystem. To achieve this, we first add an auxiliary deformation $\lambda\,y_{1}y_{2}y_{3}^{2}y_{4}$ to $\tilde{P}(\rho)$, and then we compute one period of the deformed subsystem as an analytic function in the limit of large $\lambda$ by means of the direct integration technique. In the next step, we analytically continue to the region of small $\lambda$, and finally we set $\lambda=0$. As we just explained, we first start with the residue integral expression for the period of the deformed subsystem
\begin{align}\label{directinteg}
\partial_{\xi}\Pi(\rho,\lambda)&=-\frac{1}{(1+\xi^{4})^{1/2}}\int_{\times_{i=1}^{4}
\sigma_{i}}
\frac{y_{3}\,dy_{1}dy_{2}dy_{3}dy_{4}}{\tilde{P}(\rho)-
\lambda\,y_{1}y_{2}y_{3}^{2}y_{4}}\nonumber\\
&=\frac{1}{(1+\xi^{4})^{1/2}}\sum_{n=0}^{\infty}
\frac{1}{\lambda^{n+1}}\int_{\times_{i=1}^{4}\sigma_{i}}
\frac{dy_{1}dy_{2}dy_{3}dy_{4}}{y_{1}y_{2}y_{3}y_{4}}\Big(\frac{\tilde{P}(\rho)}
{y_{1}y_{2}y_{3}^{2}y_{4}}\Big)^{n}\ .
\end{align}
In the above formula, similar to the branch I, $\sigma_{i}$ is the contour which encircles $y_{i}$ coordinate at its origin. Taking the residues with respect to $y_{i}$ coordinates in (\ref{directinteg}), we find $\partial_{\xi}\Pi$ in the limit of large $\lambda$ to be
\begin{align}\label{result1}
\partial_{\xi}\Pi(\rho,\lambda)=&\frac{1}{(1+\xi^{4})^{1/2}}\frac{1}{\lambda}\Bigg\{
\sum_{n=0}^{\infty}\sum_{k=0}^{n}\frac{(8n)!}{(8k)!(n-3k)!(n-k)!(2n-2k)!(4n-2k)!}
\frac{\rho^{8k}}{\lambda^{8n}}\nonumber\\
&+\sum_{n=0}^{\infty}\sum_{k=0}^{n}\frac{(8n+4)!}{(8k+4)!(n-3k-1)!(n-k)!(2n-2k)!
(4n-2k+1)!}\frac{\rho^{8k+4}}{\lambda^{8n+4}}\Bigg\}\ .
\end{align}
As in the previous branch, we now need to analytically continue (\ref{result1}) to the region of small $\lambda$. To perform the analytic continuation, let us focus on the first series appearing in (\ref{result1}). The second series in (\ref{result1}) has a similar structure, and as we will point out later, it does not introduce any new linearly independent solution at $\lambda=0$. It would be convenient to rewrite the fist series of (\ref{result1}) in the following way
\begin{eqnarray}\label{dPione2}
\partial_{\xi}\Pi_{1}(\rho,\lambda)=\frac{1}{(1+\xi^{4})^{1/2}}
\sum_{k=0}^{\infty}\frac{1}{(8k)!}U_{k}(\lambda)\,\rho^{8k}\ ,
\end{eqnarray}
where $U_{k}(\lambda)$ is given by
\begin{eqnarray}\label{funU}
U_{k}(\lambda)=\sum_{n=3k}^{\infty}\frac{(8n)!}{(n-3k)!(n-k)!(2n-2k)!(4n-2k)!}
\frac{1}{\lambda^{8n+1}}\ .
\end{eqnarray}
To evaluate (\ref{dPione2}) at $\lambda=0$, we first need to analytically continue (\ref{funU}) to the region of small $\lambda$. To do this, we use the standard Barnes integral representation of (\ref{funU})
\begin{eqnarray}\label{Barnes2}
U_{k}(\lambda)=\frac{1}{2\pi i}\oint_{B}d\nu\,\frac{(-1)^{\nu}\,\Gamma(-\nu)\Gamma(1+\nu)\Gamma(8\nu+1)}
{\Gamma(\nu-3k+1)\Gamma(\nu-k+1)\Gamma(2\nu-2k+1)\Gamma(4\nu-2k+1)}\,\lambda^{-8\nu-1}\ .
\end{eqnarray}
If we close the contour $B$ from the right, we recognize that (\ref{funU}) is restored by contribution of the simple poles located at $\nu=m\ ,\ (m\in{\mathbb{N}}\ ,\ m\geq3k)$. However, if we close $B$ from the left, we realize that there are four sets of simple poles on the negative side of the real axis located at $\nu=-m-a/8$ where $m=0,1,2,\cdots$, and $a\in\{1,3,5,7\}$. Similar to the previous branch, the only pole which leads to a non-vanishing result at $\lambda=0$ is located at $\nu=-1/8$. Computing the residue of this pole and plugging the result in (\ref{dPione2}), we find that at $\lambda=0$ there are eight linearly independent solutions which are given by the following hypergeometric function
\begin{eqnarray}\label{hypersol}
\partial_{\xi}\Pi(\rho)=\rho^{k}\,{ }_{5}F_{4}\left(\frac{3k+1}{24},\frac{3k+9}{24},\frac{3k+17}{24},\frac{k+1}{8},1;
\frac{k+3}{8},\frac{k+4}{8},\frac{k+7}{8},\frac{k+8}{8};\frac{2^{4}\cdot3^{3}}{8^{8}}
\rho^{8}\right)\ ,
\end{eqnarray}
where $k\in\{0,1,2,3,4,5,6,7\}$. If we analyze the second series of (\ref{result1}) in a similar manner, we find that at $\lambda=0$, the only relevant pole for analytic continuation is located at $\nu=-5/8$. Evaluating the residue of this pole, one finds exactly the eight linearly independent solutions (\ref{hypersol}).
\par
We notice that the number of solutions of the subsystem of the open/closed geometry is greater than what we expect from the Griffiths-Dwork procedure. In fact, if one applies the Griffiths-Dwork algorithm, one finds that the closure relation of the subsystem does not occur at the usual place in the diagram of the variation of Hodge structure. As a consequence, one finds for this class of examples that the order of differential operators which annihilate the periods of the subsystem is either higher or lower, depending on the deformation terms added to the Fermat polynomial. This can be also seen for the differential operators of examples studied in \cite{Aganagic:2009jq}. The relation between the Griffiths-Dwork algorithm and these higher order differential equations will be explored in a greater detail in \cite{AHJMMS}. In appendix A, we present an example in which the closure relation of the subsystem occurs at an early stage so that one is not able to compute the effective superpotential associated with the specific chosen bulk deformation.
\par
Now, let us return to solutions (\ref{hypersol}) and identify the superpotential period. According to (\ref{MFsup2}), the leading term of the superpotential is linear in  closed-string moduli. This implies that the full superpotential on this branch of the open moduli space with the chosen bulk deformations (\ref{PCY}) is given by
\begin{eqnarray}\label{Branch2sup}
{\cal W}(\psi,\phi,\xi)=\int d\xi\,\frac{\rho}{(1+\xi^{4})^{1/2}}\,{  }_{3}F_{2}(\tfrac{1}{6},\tfrac{5}{6},\tfrac{1}{4};\tfrac{5}{8},\tfrac{9}{8};
\tfrac{2^{4}\cdot3^{3}}{8^{8}}\rho^{8})\ .
\end{eqnarray}
In order to make a precise comparison with (\ref{MFsup2}), we extract the linear terms of (\ref{Branch2sup}) in terms of the two closed-string moduli $\psi$ and $\phi$. Therefore, we obtain
\begin{eqnarray}\label{linsup2}
{\cal W}^{lin}(\psi,\phi,\xi)=\psi\,\xi\,{ }_{2}F_{1}(\tfrac{1}{4},\tfrac{3}{4};\tfrac{5}{4};-\xi^{4})+\frac{1}{2}\phi\,\xi^{2}
\,{ }_{2}F_{1}(\tfrac{1}{2},\tfrac{3}{4};\tfrac{3}{2};-\xi^{4})\ .
\end{eqnarray}

In order to compare the two results, we first notice that the deformation $\psi x_1^3x_4x_2x_4$ corresponds to the contribution which comes from $s^{(3)}_{110}$, whereas the deformation $\phi x_1^3x_4x_2x_5$ corresponds to the term $s^{(3)}_{101}$. Again, the coordinates on the moduli space must be mapped correctly. The transformation here is the same as in the previous branch, (\ref{hypid}).
For the first deformation, the result of the period computation can be matched to the worldsheet result via
\eqn{
	\xi\;{}_2F_1(\tfrac{1}{4},\tfrac{3}{4};\tfrac{5}{4};-\xi^4)
	&= \frac{\Gamma(\frac{5}{4})\Gamma(\frac{1}{2})}{\Gamma(\frac{3}{4})\Gamma(1)}
	-\frac{1}{2}\frac{1}{\xi^2} \;{}_2F_1(\tfrac{3}{4},\tfrac{1}{2};\tfrac{3}{2};-\xi^{-4})
	= \text{const.}
	+\frac{\eta^4}{2}\beta^2 \;{}_2F_1(\tfrac{1}{2},\tfrac{3}{4};\tfrac{3}{2};-\beta^4) \ .
}
Clearly, this is in agreement with (\ref{MFsup2}). Similarly, for the second deformation, the chain of transformations gives rise to
\eqn{
	\frac{\xi^2}{2}\;{}_2F_1(\tfrac{1}{2},\tfrac{3}{4};\tfrac{3}{2};-\xi^4)
	&= \frac{1}{2}\frac{\Gamma(\frac{3}{2})\Gamma(\frac{1}{4})}{\Gamma(\frac{3}{4})\Gamma(1)}
	-\frac{1}{\xi} \;{}_2F_1(\tfrac{3}{4},\tfrac{1}{4};\tfrac{5}{4};-\xi^{-4})
	= \text{const.}
	-\eta^{-2} \beta \;{}_2F_1(\tfrac{1}{4},\tfrac{1}{2};\tfrac{5}{4};-\beta^4) \ ,
}
which indicates the perfect agreement between the geometric and the worldsheet computations.

\subsection{Branch III of the Open Moduli Space}
Finally, the third branch of the open moduli space is described by picking the pair $(x_3, x_4)$ which both coordinates have weight 2. Therefore, the family of holomorphic ${\mathbb{P}^{1}}$'s are embedded into the ambient projective space in the following way \eqn{\label{falcur}
	(x_1:x_2:x_3:x_4:x_5) &= (\alpha v: \beta v: u: \eta^2 u: \gamma v^2) \ .
}
The associated moduli space of the above holomorphic family of curves is a Riemann surface ${\mathbb P}^{2}_{(112)}[8]$
\eqn{
	\alpha^8+\beta^8+\gamma^4 = 0 \qquad,\qquad \eta^8=-1\ .
}
The number of holomorphic differentials on this branch is 9. The associated matrix factorization at the Fermat point is given by
\eqn{
	J_0 &= x_4-\eta^2x_3 \\
	E_0 &= x_4^3+\eta^2x_3x_4^2+\eta^4x_3^2x_4+\eta^6x_3^3 \\
	J_1 &= \alpha x_2 -\beta x_1 \\
	E_1 &= (\alpha x_2 + \beta x_1) \Bigl(\frac{x_2^6}{\alpha^2}+\frac{\beta^2 x_1^2x_2^4}{\alpha^4}+\frac{\beta^4x_1^4x_2^2}{\alpha^6}+\frac{\beta^6x_1^6}{\alpha^8}\Bigr)\\
	J_2 &= \alpha^2 x_5-\gamma x_1^2 \\
	E_2 &= \frac{x_5^3}{\alpha^2}+\frac{\gamma x_1^2x_5^2}{\alpha^4}+\frac{\gamma^2x_1^4x_5}{\alpha^6}+\frac{\gamma^3x_1^6}{\alpha^8} \ .
}
As in previous branches, we consider the complex structure deformations which have a non-vanishing correlation function. These deformations turn out to be the product of two homogeneous polynomials of degree four
\eqn{
	G &= r^{(4)}(x_3, x_4) s^{(4)}(x_1, x_2, x_5) \ .
}
Using the Kapustin-Li formula, the worldsheet bulk-boundary correlator is given by
\eqn{
	\langle G \partial_\beta Q \rangle
	=
	-\frac{1}{16}\eta^2 \alpha r^{(4)}(1,\eta^2) \frac{s^{(4)}(\alpha, \beta, \gamma)}{\gamma^3}\ .
}
This correlator can be easily integrated to an effective superpotential. Thus, for this branch of the open moduli space, we obtain
\eqn{\label{MFsup3}
	{\cal W} &= -\frac{1}{8}\eta^2  r^{(4)}(1,\eta^2)
		\sum_{abc}s^{(4)}_{abc} \frac{\eta^{10c+2}\beta^{1+b}}{b+1} {}_2F_1(\tfrac{b+1}{8}, \tfrac{3-c}{4}; \tfrac{9+b}{8}; -\beta^8)\ .
}

Let us briefly report on the results of the third branch of the open moduli space from the geometry side as well. In the current branch, the defining equation of the deformed Calabi-Yau space is given by $P=W+r^{(4)}(x_3, x_4) s^{(4)}(x_1, x_2, x_5)$. The possible candidates which can be considered for $s^{(4)}(x_{1},x_{2},x_{5})$ are limited to  $\{x_{1}^{4},x_{2}^{4},x_{1}^{3}x_{2},x_{1}x_{2}^{3},x_{1}^{2}x_{2}^{2},x_{1}^{2}x_{5},
x_{2}^{2}x_{5},x_{1}x_{2}x_{5},x_{5}^{2}\}$. Moreover, the monomials in $r^{(4)}(x_{3},x_{4})$ are chosen among $\{x_{3}^{2},x_{3}x_{4},x_{4}^{2}\}$. For simplicity of the computation, we assume that $s^{(4)}(x_{1},x_{2},x_{5})=-\psi\,x_{1}x_{2}x_{5}-\phi\,x_{1}x_{2}^{3}$ and $r^{(4)}(x_{3},x_{4})=x_{3}x_{4}$. Thus, we have
\begin{eqnarray}\label{perP}
P=W-\psi\, x_{1}x_{2}x_{3}x_{4}x_{5}-\phi\, x_{1}x_{2}^{3}x_{3}x_{4}\ .
\end{eqnarray}
In order to embed the family of holomorphic curves (\ref{falcur}) into a family of holomorphic divisors, we introduce $Q(\xi)$
\begin{eqnarray}\label{divione}
Q(\xi)=x_{5}-\xi\,x_{2}^{2}\ ,
\end{eqnarray}
where $\xi$ is the modulus of the family of divisors. Given a curve in (\ref{falcur}), the family of divisors (\ref{divione}) encompasses the curve with the value $\xi=\frac{\gamma}{\beta^{2}}$ for the open modulus. Upon the following change of coordinates
\begin{eqnarray}\label{chcoor2}
y_{1}=x_{1}\quad,\quad y_{2}=(1+\xi^{4})^{1/8}x_{2}\quad,\quad y_{3}=x_{3}\quad,\quad y_{4}=x_{4}\quad,\quad y_{5}=x_{5}-\xi\,x_{2}^{2}\ ,
\end{eqnarray}
and performing the residue integral on the $y_{5}$ coordinate, we realize that the holomorphic two-form of the subsystem of the open/closed geometry transforms to
\begin{eqnarray}\label{holtwo}
\partial_{\xi}\underline{\Omega}\simeq\frac{1}{(1+\xi^{4})^{3/8}}
\int\frac{y_{2}^{2}}{\tilde{P}(\zeta)}\Delta_{y}\ ,
\end{eqnarray}
where $\tilde{P}(\zeta)$ and the algebraic modulus $\zeta$ are defined as
\begin{eqnarray}\label{defzet}
\tilde{P}(\zeta)=y_{1}^{8}+y_{2}^{8}+y_{3}^{4}+y_{4}^{4}-\zeta\,y_{1}y_{2}^{3}y_{3}y_{4}
\qquad,\qquad \zeta=\frac{\psi\,\xi+\phi}{(1+\xi^{4})^{3/8}}\ .
\end{eqnarray}
From the definition of $\tilde{P}(\zeta)$, it is clear that the periods of the subsystem correspond to the periods of the complex surface ${\mathbb{P}}^{3}_{(1,1,2,2)}[8]$ with one algebraic modulus. Following the usual technique of direct integration, we find a period of the holomorphic two-form of the subsystem in the limit of large $\zeta$
\begin{eqnarray}\label{subper}
\partial_{\xi}\Pi(\zeta)=\frac{1}{(1+\xi^{4})^{3/8}}\sum_{n=0}^{\infty}\frac{(8n)!}
{n!\,(3n)!\,((2n)!)^{2}}\frac{1}{\zeta^{8n+1}}\ .
\end{eqnarray}
As in the previous cases, we need to analytically continue (\ref{subper}) to region of small $\zeta$. To perform this step, we use the standard Barnes integral representation of (\ref{subper})
\begin{eqnarray}\label{barnes3}
\partial_{\xi}\Pi(\zeta)=\frac{1}{(1+\xi^{4})^{3/8}}\oint_{B}d\nu\frac{(-1)^{\nu}\,
\Gamma(-\nu)\,\Gamma(8\nu+1)}{\Gamma(3\nu+1)\,\Gamma^{2}(2\nu+1)}\zeta^{-8\nu-1}\ .
\end{eqnarray}
We recognize that if we close the contour $B$ from the right, only simple poles at $\nu\in{\mathbb{N}}$ contribute, and therefore, (\ref{subper}) is restored. However, if close the contour $B$ from the left, there are six sets of simple poles that contribute the integral. These six sets of poles are located at $\nu=-k-a/8$ where $k=0,1,2,\cdots$ and $a\in\{1,2,3,5,6,7\}$. Computing the residues of these poles, we find six linearly independent periods in the regime of small $\zeta$
\begin{eqnarray}\label{LGsol}
\partial_{\xi}\Pi_{a}(\zeta)=\frac{1}{(1+\xi^{4})^{3/8}}\sum_{k=0}^{\infty}
\frac{\Gamma\big(k+\frac{a}{8}\big)\,\Gamma\big(3k+\frac{3a}{8}\big)\,\Gamma^{2}\big(2k+
\frac{a}{4}\big)}{\Gamma(8k+a)}\,\zeta^{8k+a-1}\ ,
\end{eqnarray}
where $a\in\{1,2,3,5,6,7\}$. By comparison with (\ref{MFsup3}), we figure out that the relevant period for the superpotential is $\Pi_{2}(\zeta)$. Therefore, the effective D2-brane superpotential on this branch of the open moduli space is found
\begin{eqnarray}\label{fullsup}
{\cal W}(\psi,\phi,\xi)=\int d\xi\,\frac{\zeta}{(1+\xi^{4})^{3/8}}\,{ }_{6}F_{5}\Big(\frac{1}{4},\frac{1}{4},\frac{1}{4},\frac{3}{4},\frac{7}{12},\frac{11}{12}
;\frac{3}{8},\frac{1}{2},\frac{5}{8},\frac{7}{8},\frac{9}{8};\frac{2^{4}\cdot3^{3}}{8^{8}}
\,\zeta^{8}\Big)\ .
\end{eqnarray}
Now, in order to make a precise comparison between the two superpotentials computed from the worldsheet and geometric perspectives, we extract the linear piece of (\ref{fullsup}) in terms of the closed-string moduli. We obtain
\begin{eqnarray}\label{suplin2}
{\cal W}^{lin}(\psi,\phi,\xi)=\frac{1}{4}\psi\,\xi^{2}\,{ }_{2}F_{1}(\tfrac{1}{2},\tfrac{3}{4};\tfrac{3}{2};-\xi^{4})+\frac{1}{2}\phi\,\xi\,{ }_{2}F_{1}(\tfrac{1}{4},\tfrac{3}{4};\tfrac{5}{4};-\xi^{4})\ .
\end{eqnarray}
In order to compare the geometric superpotential with the result of the worldsheet computation, we again have to perform the chain of transformations on the coordinates of the moduli space. On this branch, we have the following relation between the modulus of the family of holomorphic curves and the modulus of family of divisors
\eqn{
	-\beta^8 = \frac{1}{1+\xi^4}\ .
}
We again apply the same hypergeometric identities (\ref{hypid}) to the relevant hypergeometric functions. The first term becomes
\eqn{
	\frac{1}{4}\xi^2\; {}_2F_1(\tfrac{1}{2},\tfrac{3}{4};\tfrac{3}{2};-\xi^4)
	&= \text{const} - \frac{1}{2\xi} \; {}_2F_1(\tfrac{3}{4},\tfrac{1}{4};\tfrac{5}{4};-\xi^{-4})
	= \text{const.} - \frac{\eta^2\beta^2}{2} \; {}_2F_1(\tfrac{1}{4},\tfrac{1}{2};\tfrac{5}{4};-\beta^8)\ ,
}
while we get for the second term
\eqn{
	\frac{1}{2}\xi\; {}_2F_1(\tfrac{1}{4},\tfrac{3}{4};\tfrac{5}{4};-\xi^4)
	&= \text{const} - \frac{1}{4\xi^2} \; {}_2F_1(\tfrac{3}{4},\tfrac{1}{2};\tfrac{3}{2};-\xi^{-4})
	= \text{const.} - \frac{\eta^4\beta^4}{4} \; {}_2F_1(\tfrac{1}{2},\tfrac{3}{4};\tfrac{3}{2};-\beta^8)\ .
}
The deformation term $\psi x_1 x_2x_3x_4x_5$ corresponds to the contribution of $s^{(4)}_{111}$ to the effective superpotential, while $\phi x_1x_2^3x_3x_4$ corresponds to $s^{(4)}_{130}$. From this, we easily recognize that the effective superpotentials (\ref{MFsup3}) and (\ref{suplin2}) are identical, up to an overall constant factor.
\par
In the current section, we have considered a non-trivial example in ${\mathbb P}^{4}_{(11222)}[8]$ at the Fermat point, where the effective superpotentials have been computed in different branches of the open moduli space of a D2-branes configuration. In all cases, we have found a perfect matching between the geometric and the worldsheet results. This demonstrates the validity of the correspondence between the two approaches derived in section 3.

\section{Conclusions}

In this paper, we have analyzed the effective D5-brane superpotentials arising from type II string compactifications on Calabi-Yau manifolds, from two different perspectives. On the one hand, from the worldsheet point of view, D-branes are described by specifying certain boundary conditions for the underlying superconformal field theory defined on the worldsheet. The associated effective D-brane superpotential is then regarded as a generating functional of all disk correlation functions. On the other hand, from the geometric point of view, the effective D-brane superpotential is captured by the relative periods of the open/closed geometry. The relative periods are expressed in terms of the integral of the holomorphic three-form of the internal Calabi-Yau space over a basis of three-chains. The moduli dependence of the relative periods, and consequently of the superpotential, is governed by the variation of mixed Hodge structure of the underlying $N=1$ special geometry.
\par
In this work, we have focused on a class of models for which there exists a one dimensional open-string moduli space at the Fermat point. This flat direction corresponds to an unobstructed boundary fermion on the worldsheet. Turning on a bulk deformation, without exciting the obstructed fermions, lifts the flat direction by the effective superpotential. Having the bulk-boundary correlator at an arbitrary point in the open string moduli space, followed by an integration, yields the effective D2-brane superpotential. The bulk-boundary correlator which is the essential ingredient to compute the superpotential for this class of models is calculated by the matrix factorization techniques in the vicinity of the LG-point. By simplifying the Kapustin-Li formula which gives a handle to compute the topological CFT correlators, we have explicitly shown how the bulk-boundary correlator is mapped to the geometric superpotential which is captured by the relative periods of the corresponding open/closed geometry. More specifically, by comparison of the two approaches, we have constructed a three-chain whose volume gives rise to the effective D2-brane superpotential.
\par
Furthermore, in order to examine the general proof we presented in section 3, we have chosen an explicit example in section 4 and worked out the effective D2-brane superpotential in both worldsheet and geometric approaches. We have found that the first closed-string order superpotential computed in the geometric approach exactly coincides with the worldsheet result. To perform the computation in the geometric picture, we have employed the technique of direct integration to derive the relative periods of the open/closed system. Another way of deriving the relative periods of the open/closed geometry for this class of models is along the lines of \cite{Aganagic:2009jq} in which one first constructs the differential operators that annihilate the relative periods, and then solves the system of differential equations.
\par
It should be stressed that both the geometric and the worldsheet point of view have  some advantages and disadvantages, so that they nicely complement each other.
In particular, the effective superpotential computed from the worldsheet perspective captures the full dependence on open-string moduli, but only to first order in bulk perturbations. However, the geometric superpotential which is calculated via relative periods is exact to all orders both in closed- and open-string moduli. Therefore, by  expanding the full geometric result for the superpotential as a perturbative series in terms of the closed-string moduli, we have extracted the contribution of each closed-string order to the superpotential in a compact form. It would be very interesting to derive this result from the first principles of worldsheet obstruction techniques.
\par
On the other hand, on the conformal field theory side there can in general be additional fields that are massless to first order and have non-trivial correlation functions with other boundary or bulk fields. Since the superpotential is independent of K\"ahler moduli, all of these terms should have counterparts in the geometric regime and it would be very interesting to understand this in  detail. Some of these correlators might have a natural interpretation in the context of open/closed geometry in the same manner. In this way, one might be able to obtain a better understanding of the topological metric on the space of closed and open BRST states. This topological metric has a geometric realization in the subsystem of the open/closed geometry \cite{Alim:2009bx}.
Another line of research branching off at this point is to consider other classes of matrix factorizations, where of course one can again calculate correlators using the Kapustin-Li formula. A natural question would then be how to reproduce these results from the geometric point of view.
\par
As the last point, let us mention that the proof we have presented in this paper is centralized on the bulk perturbation around the Fermat point in moduli space. Nonetheless, the Fermat point does not play an essential role in our proof. It would be interesting to generalize this proof for the case of deformations around a rather more general point in the open/closed moduli space.

\vspace{1cm}
\noindent
{\large{\textbf{Acknowledgements}}}
\vspace{0.3cm}\\
We would like to thank Nils Carqueville, Andres Collinucci, Michael Hecht, Adrian Mertens, Daniel Plencner, and in particular Hans Jockers and Peter Mayr for fruitful discussions and comments. This work was supported by a EURYI award of the European Science Foundation.

\appendix

\section{D-brane Superpotentials and Bulk Deformations}
\setcounter{equation}{0}
As mentioned in section 4.2, from the geometric perspective, it is not always possible to find an effective D-brane superpotential by considering any arbitrary bulk deformation. To see this phenomenon in a concrete example, in this section we first choose a simple bulk deformation and show that one does not get any induced superpotential. In the second step, we investigate the root of this problem on the level of the variation of Hodge structure and show that the underlying reason for this problem is that the diagram of the variation of mixed Hodge structure is truncated at an early stage.
\par
To address this issue, let us focus on branch II of the open moduli space with the same family of holomorphic curves (\ref{famhol}). To introduce the open sector of the open/closed geometry, we embed the family (\ref{famhol}) in the same family of holomorphic divisors (\ref{div2}). Now, we would like to deform the Fermat polynomial $W$ by adding the deformation $\psi\,x_{1}^{4}x_{3}^{2}$. Thus, the deformed bulk geometry is defined by
\begin{eqnarray}\label{perW}
P(\psi)=W+\psi\,x_{1}^{4}x_{3}^{2}\ .
\end{eqnarray}
Similar to the previous cases, if we perform the following change of coordinates
\begin{eqnarray}\label{chcoorap}
y_{1}=x_{1}\quad,\quad y_{2}=x_{2}\quad,\quad y_{3}=(1+\xi^{4})^{1/2}\,x_{3}^{2}\quad,\quad y_{4}=x_{4}\quad,\quad y_{5}=x_{5}-\xi\,x_{3}\ ,
\end{eqnarray}
we easily recognize that the subsystem of the open/closed geometry is defined as a hypersurface in ${\mathbb{P}}^{3}_{(1,1,4,2)}$ with the constraint
\begin{eqnarray}\label{consap}
\tilde{P}(\rho)=y_{1}^{8}+y_{2}^{8}+y_{3}^{2}+y_{4}^{4}+\rho\,y_{1}^{4}y_{3}\qquad ,\qquad \rho=\frac{\psi}{(1+\xi^{4})^{1/2}}\ ,
\end{eqnarray}
where $\rho$ is the algebraic modulus of this complex surface. It is evident that the periods of the subsystem of the open/closed geometry defined by (\ref{consap}) coincide with the periods of the K3 surface ${\mathbb{P}}^{3}_{(1,1,4,2)}[8]$ which has one algebraic modulus. With the same technique demonstrated in section 4, we can compute the periods of the subsystem. If one does the computation, one finds
\begin{eqnarray}\label{singsol}
\partial_{\xi}\Pi(\rho)=\frac{1}{(1+\xi^{4})^{1/2}}\sum_{k=0}^{\infty}\frac{\Gamma(k+1/8)}
{k!}\Big(\frac{\rho}{2}\Big)^{2k}=\frac{\Gamma(1/8)}{(1+\xi^{4})^{1/2}}\,{ }_{1}F_{0}\Big(\frac{1}{8};\frac{\rho^{2}}{4}\Big)\ .
\end{eqnarray}
From (\ref{singsol}), it is clear that there exists only one period which is annihilated  by the set of differential operators of the subsystem. In other words, the subsystem of the open/closed geometry with the bulk deformation (\ref{perW}) possesses only one period. Now, we would like to trace this problem at the level of the variation of Hodge structure by applying the Griffiths-Dwork algorithm. In fact, we show that in this case, one finds the following truncated diagram of variation of Hodge structure.
\begin{equation}\label{Hodgediag}
\xymatrix{
F^{3}\cap W_{3} \ar[r]^{\partial_{\psi}} \ar[rd]_{\partial_{\xi}} & F^{2}\cap W_{3} \ar[d]_{\partial_{\xi}} & \\
 & F^{2}\cap W_{4/3} &
}
\end{equation}
This indicates that when all possible nonequivalent deformations are not considered, one might not be able to probe the whole middle relative cohomology group by infinitesimal variations of the complex structure moduli. To see this explicitly, let us choose the following basis for (\ref{Hodgediag})
\begin{eqnarray}\label{basis}
\underline{\pi}=(\underline{\Omega}\, ,\partial_{\psi}\underline{\Omega}\, ;\partial_{\xi}
\underline{\Omega})\ ,
\end{eqnarray}
where $\underline{\Omega}$ is the relative holomorphic three-form of the deformed Calabi-Yau. If one defines the Gauss-Manin connection on the open/closed moduli space of the open/closed geometry in the usual way
\begin{eqnarray}
\nabla_{\psi}\underline{\pi}=(\partial_{\psi}-M_{\psi})\underline{\pi}\simeq0\quad,\quad
\nabla_{\xi}\underline{\pi}=(\partial_{\xi}-M_{\xi})\underline{\pi}\simeq0\ ,
\end{eqnarray}
then one indeed finds that the closure relations of the open/closed system occur at this stage. Applying the Griffiths-Dwork procedure, the connection matrices $M_{\psi}$ and $M_{\xi}$ are explicitly given by
\begin{eqnarray}
M_{\psi}(\psi,\xi)=\left(
\begin{array}{ccc}
 0 & 1 & 0 \\
 \frac{1}{32-8 \psi ^2} & -\frac{7 \psi }{4 \left(\psi ^2-4\right)} & -\frac{\xi  \left(\xi ^4+1\right)}{2 \left(4 \xi ^4-\psi ^2+4\right) \left(\psi ^2-4\right)} \\
 0 & 0 & \frac{\psi }{16 \xi ^4-4 \psi ^2+16}
\end{array}
\right)\ ,
\end{eqnarray}
\begin{eqnarray}
M_{\xi}(\psi,\xi)=\left(
\begin{array}{ccc}
 0 & 0 & 1 \\
 0 & 0 & \frac{\psi }{16 \xi ^4-4 \psi ^2+16} \\
 0 & 0 & \frac{1}{2} \xi ^3 \left(\frac{4}{-4 \xi ^4+\psi ^2-4}-\frac{3}{\xi ^4+1}\right)
\end{array}
\right)\ .
\end{eqnarray}
These connection matrices satisfy the flatness condition, and therefore, they represent an integrable linear system of differential equations. From (\ref{basis}), it is clear that there is only one cohomology element, $\partial_{\xi}\underline{\Omega}$, corresponding to the subsystem, and hence, there exists only one period corresponding to this cohomology basis element. This explains why we do not find further solutions for this specific bulk deformation.
\par
Let us note at this point that from the matrix factorization point of view, the deformation (\ref{perW}) is not part of the perturbations considered in this paper, for reasons discussed in  \cite{Baumgartl:2007an, Baumgartl:2008qp}. The class of allowed perturbation is distinguished by the property that the bulk perturbation excites the marginal fermion generating the branch of the moduli space under consideration and no other fermion. This is not the case for the perturbation (\ref{perW}), where the correlator between the bulk deformation and the unobstructed fermion vanishes identically. However, this does not imply the existence of a new flat direction. The reason is that bulk deformations which do not correlate with $\Psi_\beta$ usually excite other marginal fermions. The other marginal fermions can, in principle, deform the matrix factorization, but since they are not exactly marginal, they lead to obstructions. This will finally lead to an effective superpotential which is generated by the full set of marginal boundary operators. In our study, we have set any modulus other than the unobstructed one to zero, and therefore, the bulk deformation  (\ref{perW}) lies outside of the class of models we have focused on here.

\bibliography{BBSref}
\bibliographystyle{hieeetr}

\end{document}